\input harvmac
\noblackbox
\ifx\answ\bigans
\magnification=1200\baselineskip=14pt plus 2pt minus 1pt
\else\baselineskip=16pt 
\fi


\def\tb{type $IIB$\ }\def\ta{type $IIA$\ }
\def\ap{\alpha'}

\def\cf{{\it cf.\ }}
\def\ie{{\it i.e.\ }}

\def\eqq{{\it Eq.\ }}
\def\eqqs{{\it Eqs.\ }}
\def\th{\theta}

\def\al{\alpha}

\def\Om{\Omega}
\def\om{\omega}

\def\Om{\Omega}

\newif\ifnref
\def\rrr#1#2{\relax\ifnref\nref#1{#2}\else\ref#1{#2}\fi}
\def\ldf#1#2{\begingroup\obeylines
\gdef#1{\rrr{#1}{#2}}\endgroup\unskip}

\def\doubref#1#2{\refs{{#1},{#2} }}
\def\threeref#1#2#3{\refs{{#1},{#2},{#3} }}
\def\fourref#1#2#3#4{\refs{{#1},{#2},{#3},{#4} }}

\def\fiveref#1#2#3#4#5{\refs{{#1},{#2},{#3},{#4},{#5} }}
\nreffalse

\def\lref{\ldf}


\def\appA{A}
\def\appB{B}

\def\tilde{\widetilde}

\def\h {{1\over 2}}

\def\ov {\overline}
\def\o {\over}
\def\fc#1#2{{#1 \o #2}}

\def\IZ{ {\bf Z}}
\def\IP{{\bf P}}\def\IC{{\bf C}}

\def\hat{\widehat}

\def\br{\hfill\break}
\def\tr {{\rm tr}}

\def\mod {{\rm mod}}
\def\lf {\left}
\def\ri {\right}
\def\ra {\rightarrow}
\def\lra {\longrightarrow}

\def\im {{\rm Im}}
\def\p {\partial}

\def\Fc {{\cal F}} 
 \def\Oc {{\cal O}}
 \def\Sc {{\cal S}}
 
 \def\Tc {{\cal T}}
 \def\Uc {{\cal U}}


\lref\LMRS{
D.~L\"ust, P.~Mayr, R.~Richter and S.~Stieberger,
``Scattering of gauge, matter, and moduli fields from intersecting branes,''
arXiv:hep-th/0404134.
}

\Title{\vbox{\rightline{HU--EP--04/32} \rightline{MPP--2004--63}
\rightline{\tt hep-th/0406092}}}
{\vbox{\centerline{Flux--induced Soft Supersymmetry Breaking in}
\bigskip\centerline{Chiral Type $IIB$ Orientifolds with $D3/D7$--Branes}}}
\smallskip
\centerline{D. L\"ust$^{a,b}$,\ \ S. Reffert$^a$, \ and\ \ S. Stieberger$^a$}
\bigskip
\centerline{\it $^a$ Institut f\"ur Physik, Humboldt Universit\"at zu Berlin,}
\centerline{\it Newtonstra\ss e 15, 12489 Berlin, FRG}
\vskip5pt
\centerline{\it $^b$ Max--Planck--Institut f\"ur Physik,}
\centerline{\it F\"ohringer Ring 6, 80805 M\"unchen, FRG}

\bigskip\bigskip
\centerline{\bf Abstract}
\vskip .2in
\noindent
We discuss supersymmetry breaking via $3$--form fluxes in chiral supersymmetric type $IIB$
orientifold vacua with  $D3$-- and $D7$--branes. 
After a general discussion of possible choices of fluxes allowing for stabilizing of 
a part of the moduli, we determine the resulting effective action including all soft supersymmetry
breaking terms. 
We also extend the computation of our previous work \LMRS\ 
concerning  the matter field metrics arising from various open string 
sectors, in particular focusing on the $1/2$ BPS $D3/D7$--brane configuration.
Afterwards, the $F$--theory lift of our constructions is investigated.

\Date{}
\noindent

\goodbreak

\lref\Wolf{O.~DeWolfe and S.B.~Giddings,
``Scales and hierarchies in warped compactifications and brane worlds,''
Phys.\ Rev.\ D {\bf 67}, 066008 (2003)
[arXiv:hep-th/0208123].
}

\lref\LustKS{
D.~L\"ust,
``Intersecting brane worlds: A path to the standard model?,''
Class.\ Quant.\ Grav.\  {\bf 21}, S1399 (2004)
[arXiv:hep-th/0401156].
}

\lref\verlinde{C.S.~Chan, P.L.~Paul and H.~Verlinde,
``A note on warped string compactification,''
Nucl.\ Phys.\ B {\bf 581}, 156 (2000)
[arXiv:hep-th/0003236].
}

\lref\greene{
B.R.~Greene, K.~Schalm and G.~Shiu,
``Warped compactifications in M and F theory,''
Nucl.\ Phys.\ B {\bf 584}, 480 (2000)
[arXiv:hep-th/0004103].
}

\lref\Bachas{
C.~Bachas,
``A Way to break supersymmetry,''
arXiv:hep-th/9503030.
}

\lref\Sagnotti{C.~Angelantonj, I.~Antoniadis, E.~Dudas and A.~Sagnotti,
``Type-I strings on magnetised orbifolds and brane transmutation,''
Phys.\ Lett.\ B {\bf 489}, 223 (2000)
[arXiv:hep-th/0007090].
}

\lref\YUK{E.~Gava, K.~S.~Narain and M.~H.~Sarmadi,
``On the bound states of p- and (p+2)-branes,''
Nucl.\ Phys.\ B {\bf 504}, 214 (1997)
[arXiv:hep-th/9704006];\br
I.~Antoniadis, K.~Benakli and A.~Laugier,
"Contact interactions in $D$--brane models,''
JHEP {\bf 0105}, 044 (2001)
[arXiv:hep-th/0011281];\br
D.~Cremades, L.~E.~Ibanez and F.~Marchesano,
``Computing Yukawa couplings from magnetized extra dimensions,''
arXiv:hep-th/0404229;
``Yukawa couplings in intersecting D-brane models,''
JHEP {\bf 0307}, 038 (2003)
[arXiv:hep-th/0302105];\br
M.~Cvetic and I.~Papadimitriou,
"Conformal field theory couplings for intersecting $D$--branes on  orientifolds,''
Phys.\ Rev.\ D {\bf 68}, 046001 (2003)
[arXiv:hep-th/0303083];\br
I.R.~Klebanov and E.~Witten,
"Proton decay in intersecting $D$--brane models,''
Nucl.\ Phys.\ B {\bf 664}, 3 (2003)
[arXiv:hep-th/0304079]\br
S.A.~Abel and A.W.~Owen,
"Interactions in intersecting brane models,''
Nucl.\ Phys.\ B {\bf 663}, 197 (2003)
[arXiv:hep-th/0303124].
}

\lref\DijkstraYM{
T.~P.~T.~Dijkstra, L.~R.~Huiszoon and A.~N.~Schellekens,
 ``Chiral supersymmetric standard model spectra from orientifolds of Gepner
arXiv:hep-th/0403196.
}

\lref\HoneckerKB{
G.~Honecker and T.~Ott,
 ``Getting just the supersymmetric standard model at intersecting branes on the
Z(6)-orientifold,''
arXiv:hep-th/0404055.
}

\lref\BlumenhagenCG{
R.~Blumenhagen and T.~Weigand,
``Chiral supersymmetric Gepner model orientifolds,''
JHEP {\bf 0402}, 041 (2004)
[arXiv:hep-th/0401148].
}

\lref\SEN{
A. Sen,
``Orientifold limit of F-theory vacua,''
Nucl.\ Phys.\ Proc.\ Suppl.\  {\bf 68}, 92 (1998)
[Nucl.\ Phys.\ Proc.\ Suppl.\  {\bf 67}, 81 (1998)]
[arXiv:hep-th/9709159];
Phys.\ Rev.\ D {\bf 55}, 7345 (1997)
[arXiv:hep-th/9702165].
}

\lref\GM{
R.~Gopakumar and S.~Mukhi,
``Orbifold and orientifold compactifications of F-theory and M-theory  to six
and four dimensions,''
Nucl.\ Phys.\ B {\bf 479}, 260 (1996)
[arXiv:hep-th/9607057].
}

\lref\GVW{S.~Gukov, C.~Vafa and E.~Witten,
``CFT's from Calabi-Yau four-folds,''
Nucl.\ Phys.\ B {\bf 584}, 69 (2000)
[Erratum-ibid.\ B {\bf 608}, 477 (2001)]
[arXiv:hep-th/9906070].
}

\lref\PM{P.~Mayr,
``Stringy world branes and exponential hierarchies,''
JHEP {\bf 0011}, 013 (2000)
[arXiv:hep-th/0006204].
}

\lref\ABFPT{I.~Antoniadis, C.~Bachas, C.~Fabre, H.~Partouche and T.R.~Taylor,
``Aspects of type I - type II - heterotic triality in four dimensions,''
Nucl.\ Phys.\ B {\bf 489}, 160 (1997)
[arXiv:hep-th/9608012].
}

\lref\CIM{
D.~Cremades, L.E.~Ibanez and F.~Marchesano,
``SUSY quivers, intersecting branes and the modest hierarchy problem,''
JHEP {\bf 0207}, 009 (2002)
[arXiv:hep-th/0201205].
}

\lref\SVW{S.~Sethi, C.~Vafa and E.~Witten,
``Constraints on low-dimensional string compactifications,''
Nucl.\ Phys.\ B {\bf 480}, 213 (1996)
[arXiv:hep-th/9606122].
}
\lref\kakui{
Z.~Kakushadze, G.~Shiu and S.H.H.~Tye,
``Type IIB orientifolds, F-theory, type I strings on orbifolds and type I
heterotic duality,''
Nucl.\ Phys.\ B {\bf 533}, 25 (1998)
[arXiv:hep-th/9804092].
}

\lref\CV{C.~Vafa,
``Evidence for F-Theory,''
Nucl.\ Phys.\ B {\bf 469}, 403 (1996)
[arXiv:hep-th/9602022].
}
\lref\BJPS{M.~Bershadsky, A.~Johansen, T.~Pantev and V.~Sadov,
``On four-dimensional compactifications of F-theory,''
Nucl.\ Phys.\ B {\bf 505}, 165 (1997)
[arXiv:hep-th/9701165].
}

\lref\DRS{K.~Dasgupta, G.~Rajesh and S.~Sethi,
``M theory, orientifolds and G-flux,''
JHEP {\bf 9908}, 023 (1999)
[arXiv:hep-th/9908088].
}

\lref\TV{T.R.~Taylor and C.~Vafa,
``RR flux on Calabi-Yau and partial supersymmetry breaking,''
Phys.\ Lett.\ B {\bf 474}, 130 (2000)
[arXiv:hep-th/9912152].
}

\lref\MayrHH{
P.~Mayr,
 ``On supersymmetry breaking in string theory and its realization in brane
worlds,''
Nucl.\ Phys.\ B {\bf 593}, 99 (2001)
[arXiv:hep-th/0003198].
}

\lref\KorsWF{
B.~K\"ors and P.~Nath,
``Effective action and soft supersymmetry breaking for intersecting D-brane
Nucl.\ Phys.\ B {\bf 681}, 77 (2004)
[arXiv:hep-th/0309167].
}

\lref\CurioSC{
G.~Curio, A.~Klemm, D.~L\"ust and S.~Theisen,
``On the vacuum structure of type II string compactifications on  Calabi-Yau
spaces with H-fluxes,''
Nucl.\ Phys.\ B {\bf 609}, 3 (2001)
[arXiv:hep-th/0012213].
}

\lref\AldazabalDG{
G.~Aldazabal, S.~Franco, L.~E.~Ibanez, R.~Rabadan and A.~M.~Uranga,
``D = 4 chiral string compactifications from intersecting branes,''
J.\ Math.\ Phys.\  {\bf 42}, 3103 (2001)
[arXiv:hep-th/0011073].
}

\lref\BL{M.~Berkooz and R.G.~Leigh,
``A D = 4 N = 1 orbifold of type I strings,''
Nucl.\ Phys.\ B {\bf 483}, 187 (1997)
[arXiv:hep-th/9605049].
}

\lref\BDL{M.~Berkooz, M.R.~Douglas and R.G.~Leigh,
"Branes intersecting at angles,''
Nucl.\ Phys.\ B {\bf 480}, 265 (1996)
[arXiv:hep-th/9606139].
}
\lref\GKP{S.B.~Giddings, S.~Kachru and J.~Polchinski,
``Hierarchies from fluxes in string compactifications,''
Phys.\ Rev.\ D {\bf 66}, 106006 (2002)
[arXiv:hep-th/0105097].
}

\lref\FP{A.R.~Frey and J.~Polchinski,
``N = 3 warped compactifications,''
Phys.\ Rev.\ D {\bf 65}, 126009 (2002)
[arXiv:hep-th/0201029].
}

\lref\uranga{M.~Cvetic, G.~Shiu and A.~M.~Uranga,
``Chiral four-dimensional N = 1 supersymmetric type IIA orientifolds from
intersecting D6-branes,''
Nucl.\ Phys.\ B {\bf 615}, 3 (2001)
[arXiv:hep-th/0107166].
}

\lref\GL{T.W.~Grimm and J.~Louis,
``The effective action of N = 1 Calabi-Yau orientifolds,''
arXiv:hep-th/0403067.
}

\lref\JGL{M.~Grana, T.W.~Grimm, H.~Jockers and J.~Louis,
``Soft supersymmetry breaking in Calabi-Yau orientifolds with D-branes and
fluxes,''
arXiv:hep-th/0312232.
}

\lref\BlumenhagenWH{
R.~Blumenhagen, L.~G\"orlich, B.~K\"ors and D.~L\"ust,
``Noncommutative compactifications of type I strings on tori with  magnetic
background flux,''
JHEP {\bf 0010}, 006 (2000)
[arXiv:hep-th/0007024].
}

\lref\CIU{P.G.~Camara, L.E.~Ibanez and A.M.~Uranga,
``Flux-induced SUSY-breaking soft terms,''
arXiv:hep-th/0311241.
}

\lref\joep{
J.~Polchinski, S.~Chaudhuri and C.V.~Johnson,
"Notes on D-Branes,''
arXiv:hep-th/9602052;\br
J.~Polchinski, 
"String Theory'', Vol. 2, Section 13, Cambridge University Press 1998.}

\lref\Carlo{C.~Angelantonj,
``Comments on open-string orbifolds with a non-vanishing B(ab),''
Nucl.\ Phys.\ B {\bf 566}, 126 (2000)
[arXiv:hep-th/9908064].
}


\lref\KST{S.~Kachru, M.B.~Schulz and S.~Trivedi,
``Moduli stabilization from fluxes in a simple IIB orientifold,''
JHEP {\bf 0310}, 007 (2003)
[arXiv:hep-th/0201028].
}

\lref\BehrndtIH{
K.~Behrndt and M.~Cvetic,
``Supersymmetric intersecting D6-branes and fluxes in massive type IIA string
theory,''
Nucl.\ Phys.\ B {\bf 676}, 149 (2004)
[arXiv:hep-th/0308045].
}

\lref\CremmerEN{
E.~Cremmer, S.~Ferrara, L.~Girardello and A.~Van Proeyen,
 ``Yang-Mills Theories With Local Supersymmetry: Lagrangian, Transformation
Nucl.\ Phys.\ B {\bf 212}, 413 (1983).
}

\lref\KKLT{S.~Kachru, R.~Kallosh, A.~Linde and S.P.~Trivedi,
``De Sitter vacua in string theory,''
Phys.\ Rev.\ D {\bf 68}, 046005 (2003)
[arXiv:hep-th/0301240].
}

\lref\LS{D.~L\"ust and S.~Stieberger,
``Gauge threshold corrections in intersecting brane world models,''
arXiv:hep-th/0302221.
}

\lref\HL{M.~Haack and J.~Louis,
``M-theory compactified on Calabi-Yau fourfolds with background flux,''
Phys.\ Lett.\ B {\bf 507}, 296 (2001)
[arXiv:hep-th/0103068].
}

\lref\BLT{R.~Blumenhagen, D.~L\"ust and T.R.~Taylor,
``Moduli stabilization in chiral type IIB orientifold models with fluxes,''
Nucl.\ Phys.\ B {\bf 663}, 319 (2003)
[arXiv:hep-th/0303016].
}

\lref\CU{J.F.G.~Cascales and A.M.~Uranga,
``Chiral 4d N = 1 string vacua with D-branes and NSNS and RR fluxes,''
JHEP {\bf 0305}, 011 (2003)
[arXiv:hep-th/0303024].
}

\lref\dewit{B. de Wit, D.J.~Smit and N.D.~Hari Dass,
``Residual Supersymmetry Of Compactified D = 10 Supergravity,''
Nucl.\ Phys.\ B {\bf 283}, 165 (1987);\br
J.M.~Maldacena and C.~Nunez,
``Supergravity description of field theories on curved manifolds and a no  go
theorem,''
Int.\ J.\ Mod.\ Phys.\ A {\bf 16}, 822 (2001)
[arXiv:hep-th/0007018].
}

\lref\LH{K. Becker, M. Becker, M. Haack and J. Louis,
``Supersymmetry breaking and $\alpha'$-corrections to flux induced  potentials,''
JHEP {\bf 0206}, 060 (2002)
[arXiv:hep-th/0204254].
}

\lref\AFT{C.~Angelantonj, S.~Ferrara and M.~Trigiante,
``New D = 4 gauged supergravities from N = 4 orientifolds with fluxes,''
JHEP {\bf 0310}, 015 (2003)
[arXiv:hep-th/0306185].
}

\lref\ADFT{C.~Angelantonj, R.~D'Auria, S.~Ferrara and M.~Trigiante,
``$K_3 \times T^2/\IZ_2$ orientifolds with fluxes, open string moduli and critical
points,''
Phys.\ Lett.\ B {\bf 583}, 331 (2004)
[arXiv:hep-th/0312019].
}

\lref\BB{K.~Becker and M.~Becker,
``M-Theory on Eight-Manifolds,''
Nucl.\ Phys.\ B {\bf 477}, 155 (1996)
[arXiv:hep-th/9605053].
}

\lref\Muduality{J.A.~Harvey, G.W.~Moore and C.~Vafa,
``Quasicrystalline Compactification,''
Nucl.\ Phys.\ B {\bf 304}, 269 (1988);\br
M.~Dine and E.~Silverstein,
``New M-theory backgrounds with frozen moduli,''
arXiv:hep-th/9712166;\br
A.~Dabholkar and J.~A.~Harvey,
``String islands,''
JHEP {\bf 9902}, 006 (1999)
[arXiv:hep-th/9809122];\br
O.J.~Ganor,
``U-duality twists and possible phase transitions in (2+1)D  supergravity,''
Nucl.\ Phys.\ B {\bf 549}, 145 (1999)
[arXiv:hep-th/9812024].
}

\lref\SOFTref{
L.E.~Ibanez and D.~L\"ust,
"Duality anomaly cancellation, minimal string unification and the effective low-energy Lagrangian 
of 4-D strings,''
Nucl.\ Phys.\ B {\bf 382}, 305 (1992)
[arXiv:hep-th/9202046];\br
V.S.~Kaplunovsky and J.~Louis,
"Model independent analysis of soft terms in effective supergravity and in string theory,''
Phys.\ Lett.\ B {\bf 306}, 269 (1993)
[arXiv:hep-th/9303040].
}
\lref\BIM{A.~Brignole, L.E.~Ibanez and C.~Munoz,
"Towards a theory of soft terms for the supersymmetric Standard Model,''
Nucl.\ Phys.\ B {\bf 422}, 125 (1994)
[Erratum-ibid.\ B {\bf 436}, 747 (1995)]
[arXiv:hep-ph/9308271].
}

\newsec{Introduction}

Type $I/II$ superstring compactifications with $D$--branes
are promising candidates to provide an effective 4-dimensional
theory very similar to the Standard Model of particle physics.
Concerning the question of how space-time supersymmetry is realized,
at least two classes of models seem to be viable scenarios.
First, space-time supersymmetry is broken by the $D$--brane
configurations. In this case the supersymmetry breaking scale is expected
to be near the string scale, and, as a consequence for solving
the hierarchy problem, large internal dimensions are most likely required.
Or, second, the open string sector from the $D$--branes preserves
N=1 supersymmetry. Then it is very natural to consider
supersymmetry breaking effects in the closed string bulk which generically transmute
themselves by gravitational interactions into the
observable open string Standard Model sector. From 
the effective field theory point of view one is dealing
with a low-energy effective N=1 supergravity theory with soft supersymmetry
breaking parameters, induced by the supersymmetry breaking effects in the bulk.
It is this second class of models which we like to study in this paper
in the context of chiral \tb orientifolds with  
supersymmetric $D3$-- and $D7$--brane configurations.

Non--trivial field strength fluxes for closed string
$p$--form fields provide  a natural
mechanism for space-time supersymmetry breaking in the bulk as well as for
stabilizing at least some of the moduli of the underlying compactification manifold.
Many attempts in this direction start from \tb superstring theory compactified on 
compact Calabi--Yau manifolds (CYM) with $3$--form $NS$ and $R$ fluxes.
An immediate consequence of allowing those non--vanishing fluxes 
is, that one in general needs to introduce extended objects in order to satisfy
the Einstein equations of the low--energy supergravity description \dewit.
The string theoretical explanation to this is, that these fluxes imply 
positive or negative $D3$--brane charge, which has to be cancelled by adding
objects producing the opposite of this charge. 
Generically, this then results in a warped form of the metric due to the back reaction
of the branes.
Typically, if the flux produces a positive $D3$--charge (so--called ISD fluxes), one may 
balance this charge by adding orientifold planes (of negative charge),  anti--$D3$--branes, or 
(wrapped) $D7$--branes with or without $2$--form fluxes.

Examples of such vacua are \tb compactified on Calabi--Yau orientifolds
(or toroidal orbifold/orientfolds) with $D3$--branes filling the uncompactified space--time.
If the $D3$ fills the uncompactified space--time, the low--energy effective action 
on the $D3$--brane is conformally invariant and described by N=4 supergravity. 
Only if the $D3$--brane sits at a singularity,
like a conifold singularity of a CYM or an orbifold singularity of an
orbifold compactification,
the gauge theory on the $D3$-branes leads to an N=1 chiral spectrum. Hence, quite generically, apart
from those special cases, the inclusion of fluxes leads to supersymmetry breaking from
N=4 to non--chiral non--supersymmetric gauge theories.
In order to obtain a chiral spectrum, one has to study more involved constructions.
{\it E.g.}\  if in addition to the $D3$--branes one has $D7$--branes with internal 
$2$--form fluxes, one may obtain chiral non--supersymmetric theories after turning
on $3$--form fluxes. 
This setup, which we want to study in more detail in this article, is $T$--dual
to \ta orientifold models with intersecting $D6$-branes
\threeref\BlumenhagenWH\AldazabalDG\uranga, which were
extensively discussed in the literature (for a review see \LustKS)\foot{Recent 
promising attempts to construct supersymmetric intersecting brane
models can be found in
\threeref\BlumenhagenCG\DijkstraYM\HoneckerKB.}$^{,}$\foot{Type $IIB$ models with
$D9/D5$--branes with magnetic $\Fc$--flux were also considered in {\it Refs.} \doubref\Bachas\Sagnotti.}.
Finally note that
\tb with $D3$-- and $D7$--branes can also be described by
$F$--theory, whose constant coupling limits describe \tb compactified on
$CY$ orientifolds.

A type $II$ compactification on a Calabi--Yau threefold $X_6$ leads to $h_{2,1}(X_6)$ 
complex structure and $h_{1,1}(X_6)$ K\"ahler moduli fields. 
Depending on how their related cohomology element 
behaves under the orientifold projection, some of the moduli fields are projected out.
The remaining fields, together with the universal dilaton field, are the scalars of N=1 chiral
multiplets. In the case of unbroken N=1 supersymmetry, and before turning
on any fluxes, these moduli fields have flat potentials and thus remain undetermined.
As we shall recall later, due to consistency, only so--called ISD--fluxes are allowed to be 
turned on at the level of lowest supergravity approximation \GKP.
After switching on such ISD--fluxes some or all of the complex structure moduli and 
the dilaton may be frozen at certain values as a result of flux quantization conditions.
On the other hand, such fluxes do not generate a scalar 
potential for the K\"ahler moduli. Even if one allowed for IASD--fluxes, the situation 
would not be improved, since the flux--dependent superpotential only depends on the dilaton and 
complex structure moduli and a potential only for those moduli
fields is generated after turning on those fluxes. Hence
those moduli remain undetermined. 
Other mechanisms, like higher loop effects, world--sheet
instanton effects or gaugino condensation were discussed to generate a potential for the K\"ahler 
moduli. 
Moreover, ISD fluxes will lead to a negative (or zero) cosmological constant.
Even after taking into account effects to stabilize the K\"ahler moduli,
the potential generically has a negative minimum. However, a small positive cosmological constant
appears to be called for by recent experimental data.
Hence, in addition to the effects mentioned above, which fix the K\"ahler moduli, other 
effects have to occur to generate a positive potential.
As suggested in \KKLT, this may be achieved by adding anti--$D3$--branes.
Our setup will be general enough to allow for these extra effects. 

If one does not want to rely on the above mentioned higher loop effects, world--sheet
instanton effects or gaugino condensation,
the introduction of $D7$--branes with 2-form fluxes in addition to the
$D3$--branes provides a natural way to fix also the K\"ahler moduli of the
internal space. 
(Moduli stabilization in type $IIB$ orientifolds with $D9$ and anti--$D9$--branes and $3$--form
fluxes
were considered in \doubref\BLT\CU, whereas \ta orientifolds with $D6$-branes and
fluxes were discussed in \BehrndtIH.)
Namely, the effective $D$--term scalar potential, which depends on the
K\"ahler moduli, is due to the attractive (or repulsive) forces between the
$D3$--branes and the $D7$--branes and also
the orientifold planes, and in its (supersymmetric) minimum, most of
the K\"ahler moduli are generically fixed. So putting together fluxes and
different
kinds of $D3/D7$--branes (or in \ta intersecting $D6$--branes), both complex
structure as well as K\"ahler moduli can be stabilized.

The main emphasis of this paper is on the computation of the soft
supersymmetry breaking
terms of the effective four--dimensional $D3/D7$ chiral gauge theory action after
turning on the $3$--form $NS$ and $R$ fluxes in the bulk.
Our derivation  of the soft terms will be performed in the framework of the $D=4$ effective
N=1 supergravity action 
\CremmerEN\
where spontaneous supersymmetry breaking is due to the non--vanishing auxiliary
$F$--term components of the moduli fields \doubref\SOFTref\BIM.
For the case of non--chiral orientifolds with only $D3$--branes, the
corresponding soft terms were already computed in 
\doubref\CIU\JGL. For intersecting $D6$--branes, a computation of
soft terms has been undertaken in  \KorsWF.
Specifically, we need the following two ingredients to compute the effective soft
terms:
First one has to determine the supersymmetry breaking $F$--terms,
which originate from an effective bulk superpotential 
\fourref\GVW\TV\MayrHH\CurioSC\ due to the
non--vanishing $3$--form fluxes, as well as the effective action
for the (closed string) moduli fields in Calabi-Yau orientifolds \GL.
Second the knowledge of the $D=4$ moduli dependent effective action of the open string
gauge and matter fields on the $D3/D7$ world volumes is required.
Computing directly the relevant string scattering amplitudes of gauge, matter and moduli
fields from intersecting $D6$--branes respective from $D9$--branes with 2-form
fluxes, the open string effective action was recently obtained
in \LMRS.
Here we will compute, via mirror symmetry and also
by direct string computations, the analogous effective action for
the open string fields on the $D3/D7$-brane system. In particular, we derive
the matter field metrics arising from various open string 
sectors, in particular focusing on the $1/2$ BPS $D3/D7$--brane configuration.
In order to deal with a specific set up, we consider an 
$\IZ_2\times\IZ_2$ orientifold with $D3$-- and $D7$--branes,
where we will derive the complete set of flux induced soft supersymmetry
breaking parameters.

The paper is organized as follows.
In the next section we introduce
type $IIB$ orientifolds with $D3$-- and $D7$--branes and $3$--form fluxes,
which are $T$--dual
to \ta orientifolds with intersecting $D6$--branes.
In section 3 we extend the results  from \LMRS\
concerning the open
string effective action to the case of $D3$--branes together
with $D7$--branes with $2$--form fluxes.
In chapter 4 we will then compute the effective action with the
$3$--form fluxes turned on, in particular the soft supersymmetry
breaking terms.
Finally, in section 5, the 
$F$--theory description of our models is provided.

\newsec{Type $IIB$ orientifolds with $D3$-- and $D7$--branes and three--form fluxes}

In this section, we shall construct \tb orientifold models with $D3$-- and
$D7$--branes, where we will allow for internal open string 2-form $f$-fluxes
turned on
the various stacks of $D7$--branes.  
Chiral fermions will arise from open strings stretched between the
$D3$--branes and the $D7$--branes, and also from open strings stretched
between
different stacks of $D7$--branes. These models can be obtained from \ta 
orientifolds
with intersecting $D6$-branes after performing $T$-duality transformations along
three internal directions. The \ta intersection angles between the
$D6$--branes correspond to the different \tb $f$-fluxes after the $T$-duality
transformations.
Specifically, we shall study the \tb superstring compactified on a six--dimensional
orbifold $X_6=T^6/\Gamma$, with the discrete group $\Gamma=\IZ_N$ or $\Gamma=\IZ_N\times\IZ_M$.
Generically, this leads to an N=2 supersymmetric (closed string) spectrum in $D=4$. 
In addition, to obtain an N=1 string spectrum
one introduces an orientifold projection $\Om I_n$, with $\Om$ describing a reversal of the 
orientation of the closed string world--sheet and $I_n$ a reflection of $n$ internal
coordinates. For $\Om I_n$ to represent a symmetry of the original theory, $n$
has to be an even integer in \tb. Moreover, in order that $\Om I_n$ becomes also a $\IZ_2$--action
in the fermionic sector, the action $\Om I_n$ has to be supplemented by the operator
$[(-1)^{F_L}]^{\lf[\fc{n}{2}\ri]}$. Here, $\lf[\fc{n}{2}\ri]$ represents the integer part of $n/2$.
The operator $(-1)^{F_L}$ assigns a $+1$ eigenvalue to states from the $NSNS$--sector and a $-1$ to
states from the $RR$--sector.
Generically, this projection produces orientifold fixed planes [$O(9-n)$--planes],
placed at the orbifold fixpoints of $T^6/I_n$. They have negative tension, which 
has to be balanced by introducing positive tension objects.
Candidates for the latter may be collections of $D(9-n)$--branes and/or non--vanishing
three--form fluxes $H_3$ and $C_3$.
In order to obtain a consistent low--energy supergravity description, the above objects
are subject to the supergravity equations of motion. Eventually, this puts restrictions on the 
possible choices of fluxes, to be discussed later.
The orbifold group $\Gamma$ mixes with the orientifold group $\Om I_n$.
As a result, if the group $\Gamma$ contains $\IZ_2$--elements $\th$, 
which leave one complex plane fixed, we obtain additional $O(9-|n-4|)$-- or $O(3+|n-2|)$--planes 
from the element $\Om I_n \th$.

Eventually, we want to
turn on vevs for the (untwisted) three--form fluxes $H_3$ and $F_3$. This limits the possibilities
for the choice of the orientifold projection $\Om I_n$. Since 
the $NS$ $2$--form $B_2$ is odd under $\Om$, we need to take $n \neq 0$ in order for
non--vanishing $3$--form flux components $H_{ijk}=\p_{[i} B_{jk]}$ 
to survive the orientifold projection.
In the case of $\Om I_6$, all $20$ real components $H_{ijk}$ may receive
non--vanishing vevs \AFT. Here, $I_6$ is the $\IZ_2$--reflection of the three internal
complex coordinates:
\eqn\reflection{
I_6\ :\ z^i\lra -z^i\ \ \ ,\ \ \ i=1,2,3\ .}

\subsec{$\IZ_2\times\IZ_2$ orientifold with $D3$-- and $D7$--branes}

As a concrete example, we concentrate on  the $\Gamma=\IZ_2\times \IZ_2$ toroidal orbifold (without
discrete torsion), with the two group generators $\th,\om$ acting in the following way
\eqn\generator{\eqalign{
\th\ :\ & (z^1,z^2,z^3)\lra (-z^1,-z^2,z^3)\ ,\cr
\om\ :\ & (z^1,z^2,z^3)\lra (z^1,-z^2,-z^3)}}
on the three internal complex coordinates $z^i\ ,\ i=1,2,3$.
Furthermore, the six--torus $T^6$ is assumed to be a direct product of three two--tori $T^{2,j}$,
\ie $T^6=\otimes_{j=1}T^{2,j}$.
The manifold $(T^2)^3/(\IZ_2\times \IZ_2)$ has the Hodge numbers $h_{(1,1)}=3$ and $h_{(2,1)}=51$.
Hence, there are three K\"ahler moduli $\Tc^j$, 
\eqn\kaehler{
\Tc^j=a^j+i\ R_1^jR_2^j\ \sin\al^j\ \ \ ,\ \ \ j=1,2,3\ ,}
describing the size of the three tori $T^{2,j}$, with the metric:
\eqn\metric{
g_j=\pmatrix{(R_1^j)^2&R_1^jR_2^j\cos \alpha^j\cr R_1^jR_2^j\cos \alpha^j&
(R_2^j)^2}\ .}
Here, the axions $a^j$ stem from reducing the $RR$ 
$4$--form on the $4$--cycles $T^{2,k}\times T^{2,l}$, \ie 
$a^j=\int\limits_{T^{2,k}\times T^{2,l}} C_4$.
Besides the three complex structure moduli
\eqn\complexs{
\Uc^j=\fc{R_2^j}{R_1^j}\ e^{i\al^j}\ ,\ \ \ j=1,2,3\ ,}
there are $48$ additional ones, which represent blowing up moduli. The latter
refer to the $3\times 16$ fixpoints resulting from the orbifold group 
elements $\th,\omega$ and $\theta\omega$. These orbifold singularities are of real codimension 4.
The respective modulus corresponds to a $C_2\times C_1$ cycle. The $C_1$ refers to
a $\IP^1$, which is collapsed at the orbifold singularity and the $C_2$ denotes
the torus, which is fixed under the respective orbifold group.
Hence we have 48 collapsed $3$--cycles $C_2\times C_1$ of type $(2,1)$ and $(1,2)$.
A detailed discussion of the massless spectrum of this model will be given in
the next section.
There is one subtlety concerning the correct field definitions.
The moduli fields $\Tc^j,\Uc^j$ we get from the geometry (we call them the
moduli in the string basis) are not scalars of chiral multiplets, so we need to
make a basis transformation to obtain the physical fields suitable for a field
theory calculation. We will now introduce the moduli $S,\,T^j,\,U^j$ in the field theory
basis. 
For the complex structure moduli, we do not need a
redefinition in \tb , $\Uc^i=U^i$. The imaginary part of the K\"ahler moduli is given
by the coupling of the gauge fields on a $D7$-brane $g^{-2}_{D7,j}$, which is wrapped on the
tori $T^{2,k}$ and $T^{2,l}$.  So 
\eqn\fieldT{
T^j=a^j+i\,{e^{-\phi_4}\over {2\pi\alpha'^{1/2}}}
\sqrt{{\im\,\Tc^k\im\,\Tc^l\over{\im\,\Tc^j}}}\ .}
The imaginary part of the dilaton $S$ is given by the gauge coupling on the
$D3$--brane $g^{-2}_{D3}$:
\eqn\fieldS{
S=C_0+i\,{e^{-\phi_4}\over {2\pi}}{\alpha'^{3/2}
\over{\sqrt{\im\,\Tc^1\im\,\Tc^2\im\,\Tc^3}}}\ .}

As described at the beginning, the orbifold group \generator\
implies  the additional orientifold actions $\Om I_6\th,\Om I_6\om$ and $\Om I_6\th\om$.
The latter essentially correspond to the generators $\Om I_2^3, \Om I_2^1$ and $\Om I_2^2$, 
respectively. The generators $I_2^j$ reflect only one complex coordinate
$z^j$: 
\eqn\reflectioni{
I_2^j\ :\ z^j\lra -z^j\ .}
The orientifold action \reflectioni\ implies 64 $O3$--planes $\Om I_6$ and 
$4\times 3=12$ $O7$--planes $\Om I_2^k\ ,\ k=1,2,3$. The latter are sitting at the
four fixed--points of each $T^{2,j}/I_2^j$.
These orientifold planes produce a negative $C_4$ and $C_8$--form potential, which
has to be cancelled.
These potentials may be balanced by placing $D3$--branes and $D7$--branes on top of the
orientifold planes. To obtain a chiral spectrum, we
may introduce (magnetic) two--form fluxes $F^j dx^j\wedge dy^j$ on the internal part of 
$D7$--brane world volume. 
Together with the internal $NS$ $B$--field\foot{Note, that $b^j$ has to be quantized due to
the orientifold projection $\Om$ to the values $b^j=0$ or $b^j=\h$ \Carlo.} $b^j$
we combine the complete $2$--form flux into 
$\Fc=\sum\limits_{j=1}^3\Fc^j:=\sum\limits_{j=1}^3(b^j+2\pi \ap F^j)\ dx^j\wedge dy^j$. 
The latter gives rise to the total internal antisymmetric background
\eqn\antisbg{
\pmatrix{0&f^j\cr-f^j&0}\ \ \ ,\ \ \ 
f^j=\fc{1}{(2\pi)^2}\ \int_{T^{2,j}} \Fc^j\ ,}
w.r.t. the $j$--th internal plane.
The $2$--form fluxes $\Fc^j$ have to obey the quantization rule:
\eqn\fluxquant{
m^j\fc{1}{(2\pi)^2\ap}\ \int_{T^{2,j}} \Fc^j=n^j\ \ \ ,\ \ \ n\in \IZ\ ,} 
\ie $f^j=\ap \fc{n^j}{m^j}$.
We obtain non--vanishing instanton numbers
\eqn\instanton{
m^j\ m^k\ \int\limits_{T^{2,j}\times T^{2,k}} \Fc\wedge \Fc =(2\pi)^4\ \ap^2\ n^j\ n^k}
on the world--volume of a $D7$--brane, which is wrapped around the $4$--cycle 
$T^{2,j}\times T^{2,k}$ with the wrapping numbers $m^j$. Hence, through the
$CS$--coupling $T_7\ C_4\wedge \Fc\wedge \Fc$, a $D7$--brane may also induce contributions to the
$4$--form potential.
Note, that a $D3$--brane may be described by a $D7$--brane with $f^j\ra\infty$.
To cancel the tadpoles arising from the Ramond forms $C_4$ and $C_8$, we introduce
$N_{D3}$ (space--time filling) $D3$--branes and $K$ stacks of $D7$--branes with internal fluxes. 
More concretely, $K^i$ stacks of $N_a^i$ $D7$--branes with internal $2$--form fluxes 
$\Fc^j,\Fc^k$ and wrapping numbers $m_a^j,m_a^k$ w.r.t. the $4$--cycle $T^{2,j}\times T^{2,k}$.
The cancellation condition for the tadpole arising from the $RR$ $4$--form $C_4$ is
\eqn\cfour{
N_{D3}+\fc{2}{(2\pi)^4\ap^2}\ \sum_{(i,j,k)\atop=\overline{(1,2,3)}}\sum_{a=1}^{K^i}\ N^i_a\ 
m_a^j\ m_a^k\ \int\limits_{T^{2,j}\times T^{2,k}} \Fc \wedge \Fc =32\ ,}
\ie according to \eqq \instanton:
\eqn\ccfour{
N_{D3}+2\  \sum_{(i,j,k)\atop=\overline{(1,2,3)}}\sum_{a=1}^{K^i}\ N^i_a\ n_a^j\ n_a^k\ =32\ .}
Furthermore, the cancellation conditions for the $8$--form tadpoles yield:
\eqn\ceight{\eqalign{
2\ \sum_{a=1}^{K^3}\ N^3_a\ m_a^1\ m_a^2&=-32\ ,\cr
2\ \sum_{a=1}^{K^2}\ N^2_a\ m_a^1\ m_a^3&=-32\ ,\cr
2\ \sum_{a=1}^{K^1}\ N^1_a\ m_a^2\ m_a^3&=-32\ .}}
The extra factor of $2$ in front of the sums over the $D7$--brane stacks accounts
for additional mirror branes. For each $D7$--brane with wrapping numbers $(m^i,m^j)$, we also
have to take into account its mirror $(-m^i,-m^j)$ in order to cancel induced
$RR$ $6$--form charges. The r.h.s. of \eqqs \cfour\ and \ceight\ accounts for the 
contributions of the $O3$-- and $O7$--planes, respectively.
An $O3$--plane contributes $-\fc{1}{4}$ of a $D3$--brane charge $T_3$. In the covering space
the $64$ $O3$--planes are doubled, thus contributing $2\times 64\times (-\fc{1}{4})=-32$ on the
l.h.s. of \ccfour.
On the other hand, in $D7$--brane charge $T_7$ units, an $O7$--plane contributes
$4T_7$. Hence, in the covering space, four $O7$--planes contribute $2\times 4\times 4=32$
on the l.h.s. of \ceight.

A $D3$--brane placed in the uncompactified $D=4$ space--time produces the
contribution
\eqn\PDthree{
V_{D3}=T_3\ e^{-\phi_4}\ \fc{\ap^{3/2}}{\sqrt{\Tc_2^1 \Tc_2^2 \Tc_2^3}}}
to the total scalar potential $V$.
Here $T_p=(2\pi)^{-p}\ap^{-\h-\fc{p}{2}}$ is the $Dp$--brane tension \joep\ and
$\phi_4=\phi_{10}-\h\ln\lf[\im(\Tc^1)\im(\Tc^2) \im(\Tc^3)/\alpha'^3\ri]$ the dilaton
field in $D=4$.
Furthermore, a $D7$--brane, wrapped around the $4$--cycle $T^{2,j}\times
T^{2,k}$
with wrapping numbers $m^j,m^k$ and internal gauge fluxes $f^k,f^l$ gives rise
to the potential
\eqn\PDseven{
V_{D7_j}=-T_7\ (2\pi)^4\ \ap^{3/2}\ e^{-\phi_4}\ m^k\ m^l\ \lf|1+i\fc{f^k}{\Tc_2^k}\ri|\
\lf|1+i\fc{f^l}{\Tc_2^l}\ri|\ \sqrt\fc{\Tc_2^k\ \Tc_2^l}{\Tc_2^j}\ .}
In order that the $D7$--branes preserve some supersymmetry, their internal $2$--form fluxes 
$f^i,f^j$  must obey the supersymmetry condition \BDL:
\eqn\susy{{f^i\over {\im\Tc^i}}=-{f^j\over{\im\Tc^j}}\ .}
In that case, the potential \PDseven\ simplifies:
\eqn\PDseveni{
V_{D7_j}=-T_7\ (2\pi)^4\ \ap^{3/2}\ e^{-\phi_4}
\ m^k\ m^l\ \lf(1-\fc{f^k\ f^l}{\Tc_2^k\ \Tc_2^l}\ri)
\ \sqrt\fc{\Tc_2^k\ \Tc_2^l}{\Tc_2^j}\ .}
Hence, the presence of $N_{D3}$ space--time filling $D3$--branes and various stacks of 
$D7$--branes produces a positive potential\foot{The extra factor of two in front of the 
$D7$--brane sum accounts for the mirror branes.}
\eqn\potentialD{
V_{D3/D7}=N_{D3}\ V_{D3}+2\ \sum_{j=1}^3 \sum_{a=1}^{K^j} N^j_a\ V_{D7_j}\ .}
Furthermore, a negative potential is generated by the presence of the
$64$\ $O3$-- and $12$ $O7_j$--orientifold planes:
\eqn\potentialDD{
V_{O3/O7}=2  e^{-\phi_4}\ \ap^{3/2}\ \lf\{-64T'_3\ \fc{1}{\sqrt{\Tc_2^1 \Tc_2^2 \Tc_2^3}}-
4T'_7\ (2\pi)^4\lf(\sqrt\fc{\Tc_2^2\ \Tc_2^3}{\Tc_2^1}+\sqrt\fc{\Tc_2^1\ \Tc_2^3}{\Tc_2^2}+
\sqrt\fc{\Tc_2^1\ \Tc_2^2}{\Tc_2^3}\ri)\ri\}.}
Here, the orientifold tension for $Op$--planes is given by $T_p'=2^{p-5}T_p$ \joep.
The extra factor of $2$ is due to the covering space.
In the case of supersymmetric $D7$--branes, \ie \susy\ holding for each brane, we have\eqn\susyi{
V_{D3/D7}+V_{O3/O7}=0\ ,}
provided the tadpole conditions \cfour\ and \ceight\ are fulfilled.

The simplest solution to the equations \cfour\ and \ceight\ is represented by the following 
example: We take $32$ space--time filling $D3$--branes and place $8$ $D7$--branes on top of each 
of the $12$ $O7$--planes. This leads to a non--chiral spectrum and the $96$ $D7$--branes 
give rise to the gauge group $SO(8)^{12}$ \GM.
A more involved example, which leads to a chiral N=1 spectrum,  can be found:
\vskip0.5cm
{\vbox{\ninepoint{
\def\ss#1{{\scriptstyle{#1}}}
$$
\vbox{\offinterlineskip\tabskip=0pt
\halign{\strut\vrule#
&~$#$~\hfil
&\vrule#
&~$#$~\hfil
&\vrule#&\vrule#
&~$#$~\hfil
&~$#$~\hfil
&~$#$~\hfil
&~$#$~\hfil
&~$#$~\hfil
&\vrule#
&~$#$~\hfil
&\vrule#
&~$#$~\hfil
&\vrule#  
\cr
\noalign{\hrule}
&
{\rm Stack}
&&
{\rm Gauge\ group}
&&&
(m^1,n^1)
&&
(m^2,n^2)
&&
(m^3,n^3)
&&
N_a
&
\cr
\noalign{\hrule}
\noalign{\hrule}
&
1
&&
U(2)
&&&
-
&&
-
&&
-
&&
4\ D3
&
\cr
&
2
&&
USp(8)
&&&
-
&&
(1,0)
&&
(-1,0)
&&
8\ D7
&
\cr
&
3
&&
USp(8)
&&&
(1,0)
&&
-
&&
(-1,0)
&&
8\ D7
&
\cr
&
4
&&
U(3)\times U(1)
&&&
(1,1)
&&
(-2,1)
&&
-
&&
8\ D7
&
\cr
&
5
&&
USp(4)
&&&
-
&&
(2,1)
&&
(-1,1)
&&
4\ D7
&
\cr
&
6
&&
U(1)
&&&
(2,1)
&&
-
&&
(-2,1)
&&
2\ D7
&
\cr
\noalign{\hrule}}}$$
\vskip-10pt
\centerline{\noindent{\bf Table 1:}
{\sl Chiral $D3/D7$ brane configuration: wrapping numbers $m^j$,}}
\centerline{\sl  internal flux numbers $n^j$ and amount of supersymmetry preserved.}
\vskip10pt}}}
\vskip-0.5cm \ \br
The supersymmetry condition \susy\ may be fulfilled for each stack, provided
the three K\"ahler moduli obey $\im T^1=\im T^3=T$ and $\im T^2=\h T$. 
Hence, in that case the $NS$--tadpoles are cancelled as well, \ie \susyi\ holds.
In addition, two of the three K\"ahler moduli $\Tc^j$ are fixed as a result
of demanding a chiral supersymmetric vacuum solution.
The full configuration preserves N=1 supersymmetry in $D=4$.
After performing $T$--dualities in the three $x$--directions of the three tori $T^{2,j}$, 
the above configuration leads to the supersymmetric intersecting $D6$--brane model, 
introduced in \uranga.

Together with the complex dilaton field
\eqn\dilaton{
S=\tau=C_0 +ie^{-\phi_{10}}\ ,}
we have $h_{(1,1)}+h_{(2,1)}+1=55$ chiral N=1 multiplets from the closed string sector (bulk):  
seven from the untwisted sector and $48$ from the twisted sector.
Additional chiral multiplets arise from the open string sector. The $D3$--branes have six
real transversal positions $\phi^i$, which combine into $3$ complex scalars. Furthermore
the $D7$--branes which are wrapped around a $4$--cycle give rise to one complex scalar
describing the transversal movement and two Wilson line moduli.
Moreover, there are  moduli accounting for bundles on the $D7$--branes.
In addition, there are moduli from the twisted (open string) sector, 
describing scalar matter fields.
We shall come to a detailed discussion of the spectrum in section 3.

\subsec{Turning on the 3-form fluxes}

Let us now give non--vanishing vevs to some of the flux components $H_{ijk}$ and $F_{ijk}$,
with $F_3=dC_2$, $H_3=dB_2$. 
A thorough discussion of $3$--form fluxes on the torus $(T^2)^3$ is presented in appendix \appA.
On the torus $T^6$, we would have 20+20
independent internal
components for $H_{ijk}$ and $F_{ijk}$. In addition, all these components remain inert
under the orientifold projection $\Om(-1)^{F_L}I_6$.
However, under
$\Om I_2$, only a subset of $12$ of the $20$ flux components survive \AFT. 
Hence, after taking into account all three $I_2$--projections $I^j_2$, we are
left with only eight flux components:
\eqn\fluxsurvive{ 
H_{135}\ , \ H_{136}\ ,\ H_{145}\ ,\ H_{146}\ ,\ H_{235}\ ,\ H_{236}\ ,\ H_{245}\ ,\ H_{246}\ .}
A similar analysis applies to the $RR$ $3$--form flux components  $F_{ijk}=\p_{[i} C_{jk]}$, 
whose $2$--form potential $C_2$ is odd under $(-1)^{F_L}$,
\ie $\Om (-1)^{F_L}\ C_2=-C_2$. With similar arguments as before, the components
\eqn\fluxsurvivei{
F_{135}\ , \ F_{136}\ ,\ F_{145}\ ,\ F_{146}\ ,\ F_{235}\ ,\ F_{236}\ ,\ F_{245}\ ,\ F_{246}}
survive the orientifold projections $\Om(-1)^{F_L}I^j_2\ ,\ j=1,2,3$. 
So far, the effect of the orbifold action $\Gamma$ has not 
yet been taken into account and the components \fluxsurvive\  and \fluxsurvivei\ 
represent the set of fluxes, invariant under both $I^j_2\ ,\ j=1,2,3$ and $I_6$.
Note, that the fluxes \fluxsurvive\ and \fluxsurvivei\ 
are automatically invariant under the orbifold group \generator\ as well and no further
components are lost. In fact, these $8$ real flux components correspond to a linear combination
of the $2h_{2,1}+2h_{3,0}=8$ primitive elements of the cohomology $H^3(X_6,\IC)$
(\cf the next subsection).
To summarize, we shall consider the \tb orbifold $(T^2)^3/(\IZ_2\times \IZ_2)$
supplemented with the additional orientifold $\Om (-1)^{F_L}I_6$ action and
obtain an N=1 (non--chiral) spectrum in the closed string sector.
After performing three $T$--duality actions in all six internal coordinates, this model
becomes the non--chiral $D9/D5$--orientifold model of \BL.

The fluxes 
in \fluxsurvive\ and \fluxsurvivei\
have to obey the quantization rules
$\fc{1}{(2\pi)^2\ap}\int_{C_3} F_3\in\IZ$ and $\fc{1}{(2\pi)^2\ap}\int_{C_3} H_3\in\IZ$,
with $F_3, H_3 \in H^3(X_6,\IZ)$. 
We shall make some comments in the following.
It has been pointed out in {\it Ref.} \FP, that there are subtleties for toroidal orientifolds  
due to additional $3$--cycles, which are not present in the covering space $T^6$.
If some integers are odd, additional discrete flux has to be
turned on in order to meet the quantization rule for those $3$--cycles.
We may bypass these problems in our concrete $\IZ_2\times \IZ_2$ orientifold, 
if we choose the quantization numbers to be multiples of $8$ 
and do not allow for discrete flux at the orientifold planes \doubref\BLT\CU.
Note, that in addition to the untwisted flux components $H_{ijk}$ and $F_{ijk}$ there may
be also $NSNS$-- and $RR$--flux components from the twisted sector. We do not
consider them here. It is assumed, that their quantization rules freeze the blowing up 
moduli at the orbifold singularities. The extra $48$ $3$--cycles, which are collapsed
at those singularities, do not give rise to extra quantization conditions for
the untwisted flux components. In fact, the flux integrals over those give zero \DRS.

The two $3$--forms $F_3,H_3$ are organized in the $SL(2,\IZ)_S$ covariant field:
\eqn\fluxcomb{
G_3=F_3-S H_3\ .}
After giving a vev to the field $G_3$, the Chern--Simons 
term\foot{Throughout
this section, we work in the string--frame, \ie with the Einstein term
$\fc{1}{(2\pi)^7\ \alpha'^4}\ \int d^{10}x\ \sqrt{-g_{10}}\
e^{-2\phi_{10}}\  R$.}
\eqn\CSbulk{
\Sc_{CS}=\h\ \fc{1}{(2\pi)^7\ \ap^4}\ \int \fc{C_4\wedge
G_3\wedge\ov G_3}{S-\ov S}\ .}
of the 
ten--dimensional effective \tb action gives rise to an additional tadpole for the 
$RR$ four--form $C_4$ (in units of $T_3$):
\eqn\Nflux{
N_{flux}=\fc{1}{(2\pi)^{4}\ \ap^{2}}\ \int_{X_6} H_3\wedge F_3\ .}
Hence in the presence of $3$--form fluxes, the tadpole condition \cfour\ is modified to:
\eqn\modcfour{
N_{flux}+N_{D3}+
\fc{2}{(2\pi)^4\ap^2}\ \sum_{(i,j,k)\atop=\overline{(1,2,3)}}\sum_{a=1}^{K^i}\ N^i_a\ 
m_a^j\ m_a^k\ \int\limits_{T^{2,j}\times T^{2,k}} \Fc\wedge \Fc =32\ .}
The $CP$ even analog of \Nflux\ originates from the piece 
$\fc{-1}{2\cdot 3!}\fc{1}{(2\pi)^7\ap^4}
\int d^{10}x\ \sqrt{-g_{10}}\ |G_3|^2$ of the $D=10$ \tb action and leads to
the potential term in $D=4$:
\eqn\CPeven{
\fc{1}{2\ (2\pi)^7\ \ap^4}\ \int_{X_6} d^6y\ G_3\wedge\star_6\ov G_3\ .}
According to \doubref\GKP\KST, the latter may be split into a purely topological term $V_{top}$, 
independent of the moduli fields, and a second term $V_{flux}$, relevant for the
$F$--term contribution to the scalar potential.
After the decomposition $G_3=G^{ISD}+G^{IASD}$, with
$\star_6 G^{ISD}=+i\ G^{ISD}$ and $\star_6 G^{IASD}=-i\ G^{IASD}$, one obtains \threeref\GKP\KST\Wolf:
\eqn\fluxdec{\eqalign{
V_{flux}&=\fc{1}{(2\pi)^7\ \ap^4}\ \int_{X_6}\ G^{IASD}\wedge \star_6\ov G^{IASD}
\ ,\cr
V_{top}&=-e^{-\phi_{10}}\ T_3\ N_{flux}\ .}}
Hence, the total contributions to the scalar potential are:
\eqn\total{\eqalign{
V&=V_D+V_F\ ,\cr
V_D&=V_{D3/D7}+V_{O3/O7}+V_{top}\ ,\cr
V_F&=V_{flux}\ .}}
The piece $V_D$ represents $D$--term contributions to the scalar potential
due to Fayet--Iliopolous terms. See {\it Ref.} \CIM\ for further details.
Only the last term corresponds to an $F$--term.
For the case, that the conditions \modcfour\ and \ceight\ are met, Ramond tadpole contributions 
are absent. If in addition, \susy\ is met, \ie only supersymmetric $2$--form
fluxes on the $D7$--brane world--volume are considered, the first three terms add up to zero: 
$V_{D3/D7}+V_{O3/O7}+V_{top}=0$, \ie $V_D=0$. 
Let us remark, that this condition may generically also fix some of the
K\"ahler moduli $\Tc^j$.
In the following, we shall assume, that $V_D=0$ and study only the
$F$--term contribution $V_F=V_{flux}$ to the scalar potential $V$.
The potential $V_F$, displayed in \eqq \total, originates from the closed string sector only.
It is derived from the superpotential\foot{The factor of $\lambda$ serves to obtain the 
correct mass dimension of 3 for the superpotential, i.e. $\lambda \propto \kappa_4^{-3}$} \TV:
\eqn\TVW{
\hat W={\lambda\over{(2\pi)^2\alpha'}}\int_{X_6} G_3\wedge \Omega\ .}

\subsec{The 3-form fluxes on $(T^2)^3/\IZ_2\times \IZ_2$}

We will now work out the explicit form of the 3-form flux $G_3$ in terms of
the closed string moduli $U^i$ and $S$. 
Expressed in real coordinates, we take the following basis of 3-form fluxes
allowed on $(T^2)^3/\IZ_2\times \IZ_2$:
\eqn\realbase{
\eqalign{\alpha_0&=dx^1 \wedge dx^2  \wedge dx^3 \qquad \beta^0=dy^1 \wedge dy^2 \wedge dy^3 \cr
\alpha_1&=dy^1 \wedge dx^2  \wedge dx^3 \qquad  \beta^1=-dx^1 \wedge dy^2
 \wedge dy^3 \cr
\alpha_2&=dx^1 \wedge dy^2  \wedge dx^3  \qquad  \beta^2=-dy^1 \wedge dx^2 \wedge dy^3 \cr
\alpha_3&=dx^1 \wedge dx^2  \wedge dy^3  \qquad  \beta^3=-dy^1 \wedge dy^2 \wedge dx^3 \cr}}
This basis has the property $\int_{X_6} \alpha_i \wedge \beta^j=\delta^j_i$.
The above fluxes all fulfill the primitivity condition $\alpha_i\wedge J=\beta^i\wedge J=0$, with
$J$ the K\"ahler form. These are at the same time exactly the fluxes that are
allowed in a setup with three stacks of $D7$ branes, one
stack not wrapping $T_3^2$, one not wrapping  $T_2^2$, and one not wrapping
$T_1^2$, see appendix 1.

Expressed in this basis, 
the $G_3$-flux \fluxcomb\ takes the following form:
\eqn\GR{{1\over{(2\pi)^2\alpha'}}{G_3}=\sum_{i=0}^{3}\{(a^i-S
c^i)\alpha_i+(b_i-S d_i)\beta^i\}.}
A basis of $H^3=H^{(3,0)}\oplus H^{(2,1)}\oplus H^{(1,2)}\oplus H^{(0,3)}$,
corresponding to the fluxes \realbase \ 
is
\eqn\cplxz{\eqalign{
&dz^1\wedge dz^2\wedge dz^3,\ d\ov z^1\wedge dz^2\wedge dz^3,\ dz^1\wedge
d\ov z^2\wedge dz^3,\ dz^1\wedge dz^2\wedge d\ov z^3,\cr
& dz^1\wedge d\ov z^2\wedge d\ov z^3,\
d\ov z^1\wedge dz^2\wedge d\ov z^3,\ d\ov z^1\wedge d\ov z^2\wedge dz^3,\
d\ov z^1\wedge d\ov z^2\wedge d\ov z^3,}}
where $dz^i=dx^i+U^idy^i$.
We now want to express this basis through our real basis $\{\alpha_i, \
\beta^j\}$ and the complex structure moduli of  $T^2 \times T^2 \times T^2$ :
\eqn\cplxbase{
\eqalign{
\omega_{A0}&= \alpha_0+ \sum_{i=1}^3 \alpha_i\ov{U^i} -
\beta_1\ov{U^2}\ov{U^3} - \beta_2\ov{U^1}\ov{U^3} -
\beta_3\ov{U^1}\ov{U^2} + \beta_0\ov{U^1}\ov{U^2}\ov{U^3} \cr 
\omega_{A1}& = \alpha_0 +\alpha_1\ov{U^1}+\alpha_2U^{2}+\alpha_3U^3 - \beta_1U^2U^3 -
\beta_2\ov{U^1}U^3 - \beta_3\ov{U^1}U^2 + \beta_0\ov{U^1}U^2U^3 \cr
\omega_{A2}& = \alpha_0 +\alpha_1U^1+\alpha_2\ov{U^2}+\alpha_3U^3  - \beta_1\ov{U^2}U^3 -
\beta_2U^{1}U^3 - \beta_3U^{1}\ov{U^2} + \beta_0U^{1}\ov{U^2}U^3 \cr
\omega_{A3}& = \alpha_0 +\alpha_1U^1+\alpha_2U^{2}+\alpha_3\ov{U^3}  - \beta_1U^{2}\ov{U^3} -
\beta_2U^{1}\ov{U^3} - \beta_3U^{1}U^{2} + \beta_0U^{1}U^{2}\ov{U^3} \cr
\omega_{B0}&=\alpha_0+ \sum_{i=1}^3 \alpha_iU^i -
\beta_1U^2U^3 - \beta_2U^1U^3 - \beta_3U^1U^2 + \beta_0U^1U^2U^3\cr
\omega_{B1}& = \alpha_0 +\alpha_1U^{1}+\alpha_2\ov{U^2}+\alpha_3\ov{U^3} -
\beta_1\ov{U^2}\ov{U^3} -
\beta_2U^{1}\ov{U^3} - \beta_3U^{1}\ov{U^2} + \beta_0U^{1}\ov{U^2}\ov{U^3} \cr
\omega_{B2}& = \alpha_0 +\alpha_1\ov{U^1}+\alpha_2U^{2}+\alpha_3\ov{U^3}  - 
\beta_1U^{2}\ov{U^3} -
\beta_2\ov{U^1}\ov{U^3} - \beta_3\ov{U^1}U^{2} + \beta_0\ov{U^1}U^2\ov{U^3} \cr
\omega_{B3}& = \alpha_0 +\alpha_1\ov{U^1}+\alpha_2\ov{U^2}+\alpha_3U^3  - \beta_1\ov{U^2}U^{3} -
\beta_2\ov{U^1}U^{3} - \beta_3\ov{U^1}\ov{U^2} +
\beta_0\ov{U^1}\ov{U^2}U^{3} \cr}}
 $\omega_{B0}$
obviously corresponds to the $(3,0)$-part of the flux and the Calabi-Yau 3-form
$\Omega$ can be normalized to equal  $\omega_{B0}$.  $\omega_{A1}$,
$\omega_{A2}$ and  $\omega_{A3}$ are the $(2,1)$-components of the flux,
$\omega_{B1}$, $\omega_{B2}$ and $\omega_{B3}$ the $(1,2)$-components of the
flux, and $\omega_{A0}$ corresponds to the $(0, 3)$-part, \ie~$\ov{\Omega}$.

Note that this basis is not normalized to one. It fulfills
\eqn\cmplxbaserel{\eqalign
{
\int\om_{Ai}\wedge\om_{Bi}&=\prod_{i=1}^3(U^i-\ov U^i),\quad i=0,\ldots,3\cr
\int\om_{Aj}\wedge\om_{Bk}&=0,\quad j\neq k.}}
Expressed in this basis,  the $G_3$-flux takes the following form:
\eqn\GC{{1\over{(2\pi)^2\alpha'}}G_3=\sum_{i=0}^3 (A^i\omega_{Ai}+B^i\omega_{Bi}).}
By comparing the coefficients of $G_3$ expressed in the real basis and in the
complex basis and solving for the $\{A^i,\ B^i\}$, we can express the $\{A^i,\ B^i\}$ as a
function of $\{a^i, c^i, b_i, d_i\}$ and the moduli fields $S, U^i$.
By setting the respective coefficients to zero, we obtain equations for the
respective flux parts. This gives us constraints on the $\{a^i,
c^i, b_i, d_i\}$. What has to be taken into account as well is the fact that
the $\{a^i, c^i, b_i, d_i\}$ must be integer numbers. This requirement can
only be fulfilled for specific choices of the $U^i$ and of $S$, i.e. it
fixes the moduli.

Expressed with the coefficients of the real basis, we find $N_{\rm flux}$,
given in \Nflux, to be
$$N_{\rm flux}=\sum_{i=0}^3 c^ib_i-\sum_{i=0}^3a^id_i.$$
We want to find the corresponding expression in complex language. We find 
\eqn\cplxNflux{
\eqalign{
N_{\rm flux}&={1\over {(2\pi)^4(\alpha')^2}}{1\over {(S-\ov S)}}\int\ov
G_3\wedge G_3\cr
&=-{{\prod\limits_{i=1}^3(U^i-\ov U^i)}\over{(S-\ov
S)}}\sum_{i=0}^3(|A^i|^2-|B^i|^2),
}}
which is quite a nice expression. And it immediately teaches us something about
the behaviour of the different fluxes: The fluxes obeying the ISD-condition,
i.e. those having all $B_i=0$, have $N_{\rm flux}>0$, whereas the IASD-fluxes,
i.e. those with all $A_i=0$ have $N_{\rm flux}<0$.

\vskip0.2cm
\noindent
{\it (i) The supersymmetric case: $(2,1)-$flux}
\vskip0.2cm

It is common knowledge that turning on only the $(2,1)$-part of the 3-form
flux does not break supersymmetry. Such a flux fulfills the ISD condition
and from eq. \fluxdec\ , we know that $V_{\rm flux}=0$.

We obtain the necessary equations by
setting the coefficients of the $(0,3)$-, $(1,2)$-, and $(3,0)$-flux to zero,
i.e. $A_0=B_0=B_1=B_2=B_3=0$. This corresponds to the following equations:
\eqn\Eq{\eqalign{
0&= U^1U^2U^3(a^0-S c_0)-\sum_{i\neq j\neq k}(a^i-S c^i)U^jU^k-(b_0-S d_0)-
\sum_{i=1}^3(b_i-S d_i) U^i\cr
0&= \ov{ U}^1U^2U^3(a^0-S c_0)-\{(a^1-Sc^1)U^2U^3+(a^2-S c^2)
\ov{U}^1U^3+(a^3-S c^3) \ov{U}^1U^2\}-\cr
&-(b_0-Sd_0)-\{(b_1-S d_1) \ov{ U}^1+(b_2-S d_2) U^2+(b_3-S d_3) U^3\} \cr
0& =\ov{U}^1\ov{U}^2\ov{U}^3(a^0-S c_0)-\sum_{i\neq j\neq k}(a^i-S c^i)\ov{U}^j\ov{U}^k-(b_0-S
d_0)-\sum_{i=1}^3(b_i-S d_i)\ov{ U}^i \cr
0&= U^1 \ov{U}^2U^3(a^0-S c_0)-\{(a^1-S c^1)
\ov{U}^2U^3+(a^2-S c^2)U^1U^3+(a^3-S c^3)U^1 \ov{U}^2\}-\cr
&-(b_0-S
d_0)-\{(b_1-S d_1) U^1+(b_2-S d_2) \ov{ U}^2+(b_3-S d_3) U^3\} \cr
0&= U^1U^2 \ov{U}^3(a^0-S c_0)-\{(a^1-S c^1)U^2
\ov{U}^3+(a^2-S c^2)U^1 \ov{U}^3+(a^3-S c^3)U^1U^2\}-\cr
&-(b_0-S
d_0)-\{(b_1-S d_1) U^1+(b_2-S d_2) U^2+(b_3-S d_3)  \ov{U}^3\},
}}
with $i\neq j\neq k$.
There is another, more elegant way of obtaining the equations for the
$(2,1)$-flux: 
$G_3$ not having a $(0,3)$-part is equivalent to requiring that
$$\int G_3\wedge \Omega =0,$$
which yields the first of the above equations. $G_3$ not having a $(3,0)$-part is equivalent 
to requiring that
$$\int G_3\wedge \ov{\Omega} =0,$$ which yields the second of the above equations,
and $G_3$ not having a $(1,2)$-part is equivalent to requiring that
$$\int G_3\wedge \omega_{A1}=\int G_3\wedge \omega_{A2}=\int
G_3\wedge\omega_{A3}=0,$$ which gives us the remaining three equations.
(We remember that $\omega_{A1},\ \omega_{A2},\ \omega_{A3}$ are the basis of
the $(2,1)$-flux.) The integration picks out the allowed flux components and
we obtain the same equations as above in a different way.

There is another equivalent  way to obtain the equations for the $(2,1)$-flux. We know
that for the flux to be supersymmetric, we must impose the conditions
\eqn\susycond{\eqalign
{\hat W&={\lambda\over{(2\pi)^2\alpha'}}\ \int G_3\wedge \Om=0,\cr
D_S\hat W&=\partial_{S}\hat W+\kappa_4^2\ \hat W\ \partial_{S}\hat K=0,\cr
D_{U^i}\hat W&=\partial_{U^i}\hat W+\kappa_4^2\ \hat W\ \partial_{U^i}\hat K=0,}}
where $D_M$ is the K\"ahler covariant derivative and the K\"ahler potential $\hat K$ is given in the
next section.
We want to check, whether these conditions are really equivalent to the ones
we found before. The first condition obviously corresponds to our first equation. After multiplication of
the second condition with $-(S-\ov S)$, we find it to correspond to $\int\ov
G_3\wedge\Om=0$, which is the complex conjugate of the second of our
equations. After multiplicating the third of the conditions with $-(U^i-\ov
U^i)$, we find it to correspond to $\int G_3\wedge \omega_{Ai}$, so we have
complete equivalence between our equations and the conditions above.

Now we can solve for the $\{a^i, c^i,b_i, d_i\}$ and impose the constraint
that they have to be integer numbers. These constraints cannot be solved in
full generality, i.e. for arbitrary moduli and flux coefficients. By fixing
some of the moduli and/or flux coefficients, it is possible to obtain special
solutions. Here, we choose to fix the moduli to $U^1=U^2=U^3=S=i$.

One possible solution for the $(2,1)$-flux we
get is

\eqn\eqzweieins{\eqalign{
{1\over{(2\pi)^2\alpha'}}\ G_{21}&=[-d_0+i(d_1+d_2+d_3)]\ \alpha_0+[-d_1-i(-b_2-b_3+d_0)]\ 
\alpha_1\cr
&+(-d_2-ib_2)\ \alpha_2+(-d_3-ib_3)\ \alpha_3+(-d_1-d_2-d_3-id_0)\ \beta^0\cr
&+(-b_2-b_3+d_0-id_1)\ \beta^1+(b_2-id_2)\ \beta^2+(b_3-id_3)\ \beta^3,}}
where $b_2, b_3, d_0, d_1, d_2, d_3$ can be any integer number. As reported
in section 2, we can avoid possible complications with flux quantization, if
we take our flux coefficients to be multiples of 8. This can be
achieved by simply taking  $b_2, b_3, d_0, d_1, d_2, d_3$ to be multiples of
8.

Expressed in the complex basis \cplxbase, the solution takes the form
\eqn\eqzweieinscplx{\eqalign{
{1\over{(2\pi)^2\alpha'}}\ G_{21}&=\half\ [-b_2-b_3+i(d_2+d_3)]\ \om_{A1}+
\half\ [b_2-d_0+i(d_1+d_3)]\ \om_{A2}+\cr
&+\half\ [b_3-d_0+i(d_1+d_2)]\ \om_{A3}\ .}}
For $N_{\rm flux}$ we find
\eqn\Nfluxsusy{\eqalign{
N_{\rm flux}&=4\ (|A^1|^2+|A^2|^2+|A^3|^2)\cr
&=2\ (\sum_{i=0}^3d_i^2+d_1d_2+d_1d_3+d_2d_3+b_2^2+b_3^2+b_2b_3-b_2d_0-b_3d_0)\ .}}
If we require $N_{\rm flux}$ to have a certain value, this places quite
stringent constraints on our choice for the coefficients. The smallest
possible $N_{\rm flux}$ for our solution, the coefficients being multiples of
8, is $N_{\rm flux}=128$. To achieve this, we have several possibilities. We can
for example set either of the $d_i$ or $b_i$ to $\pm 8$, and all the other coefficients to
zero. For $d_0=8$ for example, all other coefficients being zero, this would amount to
$${1\over{(2\pi)^2\alpha'}}G_3=8\ (-\alpha_0-i\alpha_1-i\beta^0+\beta^1).$$
Another possible solution would be $b_2=8,\quad b_3=-8$, or the other way
round. This would result in
$${1\over{(2\pi)^2\alpha'}}G_3=8\ (-i\alpha_2+i\alpha_3+\beta^2-\beta^3).$$

\vskip0.2cm
\noindent
{\it (ii) $(0,3)$-flux}
\vskip0.2cm

This flux meets the ISD-condition as well, therefore $V_{\rm flux}=0$. 

To obtain the $(0,3)$-part of the flux, we must set
$A_1=A_2=A_3=B_0=B_1=B_2=B_3=0$ or equivalently require that $\int G_3\wedge
\ov{\Omega}=\int G_3\wedge \omega_{A1}=\int G_3\wedge \omega_{A2}=\int
G_3\wedge\omega_{A3}=\int G_3\wedge \omega_{B1}=\int G_3\wedge
\omega_{B2}=\int G_3\wedge\omega_{B3}=0$. This results in the following seven
equations:
\eqn\Eq{\eqalign{
0&= U^1\ov{U}^2\ov{U}^3(a^0-S c_0)-\{(a^1-S
c^1)\ov{U}^2\ov{U}^3+(a^2-S c^2) \ov{U}^1\ov{U}^3+(a^3-S c^3)
U^1\ov{U}^2\}-\cr
&-(b_0-S
d_0)-\{(b_1-S d_1)U^1+(b_2-S d_2) \ov{U}^2+(b_3-S d_3)\ov{ U}^3\}
\cr
0&= \ov{U}^1 U^2\ov{U}^3(a^0-S c_0)-\{(a^1-S c^1)U^2\ov{U}^3+(a^2-S c^2)
\ov{U}^1\ov{U}^3+(a^3-S c^3)\ov{U}^1U^2\}-\cr
&-(b_0-S
d_0)-\{(b_1-S d_1) \ov{U}^1+(b_2-S d_2)U^2+(b_3-S d_3) \ov{U}^3\} \cr
0&= \ov{U}^1\ov{U}^2 U^3(a^0-S c_0)-\{(a^1-S c^1)\ov{U}^2
U^3+(a^2-S c^2)\ov{U}^1U^3+(a^3-S c^3)\ov{U}^1\ov{U}^2\}-\cr
&-(b_0-S
d_0)-\{(b_1-S d_1) \ov{U}^1+(b_2-S d_2)\ov{ U}^2+(b_3-S d_3)U^3\}\cr
0&= \ov{U}^1\ov{U}^2\ov{U}^3(a^0-S c_0)-\sum_{i\neq j\neq k}(a^i-S c^i)\ov{U}^j
\ov{U}^k-(b_0-\tau
d_0)-\sum_{i=1}^3(b_i-S d_i) \ov{U}^i\cr
0&= \ov{ U}^1U^2U^3(a^0-S c_0)-\{(a^1-S
c^1)U^2U^3+(a^2-S c^2) \ov{U}^1U^3+(a^3-S c^3) \ov{U}^1U^2\}-\cr
&-(b_0-S
d_0)-\{(b_1-S d_1) \ov{ U}^1+(b_2-S d_2) U^2+(b_3-S d_3) U^3\} \cr
0&= U^1 \ov{U}^2U^3(a^0-S c_0)-\{(a^1-S c^1)
\ov{U}^2U^3+(a^2-S c^2)U^1U^3+(a^3-S c^3)U^1 \ov{U}^2\}-\cr
&-(b_0-S
d_0)-\{(b_1-S d_1) U^1+(b_2-S d_2) \ov{ U}^2+(b_3-S d_3) U^3\} \cr
0&= U^1U^2 \ov{U}^3(a^0-S c_0)-\{(a^1-S c^1)U^2
\ov{U}^3+(a^2-S c^2)U^1 \ov{U}^3+(a^3-S c^3)U^1U^2\}-\cr
&-(b_0-S d_0)-\{(b_1-S d_1) U^1+(b_2-S d_2) U^2+(b_3-S d_3)\ov{U}^3\},\cr
}}
with $i\neq j\neq k$.
Now we solve again for the $\{a^i, c^i,b_i, d_i\}$, and after fixing the moduli
$U^1=U^2=U^3=S=i$ get a solution for the $(0,3)$-flux:
\eqn\eqnulldrei{\eqalign{
{1\over{(2\pi)^2\alpha'}}G_{03}&=(d_0+id_3)\alpha_0+(d_3-id_0)\alpha_1\cr
&+(d_3-id_0)\alpha_2+(d_3-id_0)\alpha_3-(d_3-id_0)\beta^0\cr
&+(d_0+id_3)\beta^1+(d_0+id_3)\beta^2+(d_0+id_3)\beta^3.}}
We see, that we now have much stronger constraints than in the $(2,1)$-case,
which is no surprise, as we also have more equations to fulfill. When we could
choose any integer numbers for $b_2, b_3, d_0, d_1, d_2, d_3$ in the
$(2,1)$-case, we can now only choose the values for $d_0$ and $d_3$, which we
again take to be multiples of 8.
Expressed in the complex basis \cplxbase, the solution takes the form
\eqn\eqnulldreicplx{\eqalign{
{1\over{(2\pi)^2\alpha'}}G_{03}&=(d_0+id_3)\om_{A0}.}}
For the $(0,3)$-flux, we find
$$N_{\rm flux}=4\ |A_0|^2=4\ (d_3^2+d_0^2).$$
Here, the smallest possible $N_{\rm flux}$ is $256$. There are four possible
solutions for $N_{\rm flux}=256$, namely $d_3=\pm 8,\ 
d_0=0$ and  $d_0=\pm 8,\ d_3=0$. For $d_3=8,\ d_0=0$, this results in
$${1\over{(2\pi)^2\alpha'}}G_3=8\ (i\alpha_0+\alpha_1+\alpha_2+\alpha_3-\beta^0+i\beta^1+
i\beta^2+i\beta^3).$$

\vskip0.2cm
\noindent
{\it (iii) $(1,2)$-flux}
\vskip0.2cm

This flux is IASD, and therefore not consistent with the supergravity equations
of motion.

Here, we require $A_0=A_1=A_2=A_3=B_0=0\ $ or equivalently $\int G_3\wedge\Omega=\int G_3\wedge
\ov{\Omega}=\int G_3\wedge \omega_{B1}=\int G_3\wedge
\omega_{B2}=\int G_3\wedge\omega_{B3}=0$. We will refrain from writing down
the equations for this case, which are just a different combination of the
equations we have seen before.
A solution for $U^1=U^2=U^3=S=i$ is
\eqn\eqeinszwei{\eqalign{
{1\over{(2\pi)^2\alpha'}}G_{12}&=[d_0+i(d_1+d_2+d_3)]\ \alpha_0+[d_1-i(b_2+b_3+d_0)]\ \alpha_1\cr
&+(d_2+ib_2)\ \alpha_2+(d_3+ib_3)\ \alpha_3+[d_1+d_2+d_3-id_0]\ \beta^0\cr
&+(-b_2-b_3-d_0-id_1)\ \beta^1+(b_2-id_2)\beta^2+(b_3-id_3)\ \beta^3,}}
which differs from the $(2,1)$-case only by signs.
Expressed in the complex basis \cplxbase, the solution takes the form
\eqn\eqzeinszweicplx{\eqalign{
{1\over{(2\pi)^2\alpha'}}G_{12}&=\half[-b_2-b_3+i(d_2+d_3)]\ \om_{B_1}+\half[b_2+d_0+i(d_1+d_3)]
\ \om_{B_2}+\cr
&+\half[b_3+d_0+i(d_1+d_2)]\ \om_{B3}.}}
For
$(1,2)$-flux, we find
\eqn\Nfluxeinszwei{\eqalign{
N_{\rm flux}&=-4\ (|B^1|^2+|B^2|^2+|B^3|^2)\cr
&=-2\ (\ \sum_{i=0}^3d_i^2+d_1d_2+d_1d_3+d_2d_3+b_2b_3+b_2d_0+b_3d_0+b_2^2+b_3^2\ ).}}
As remarked before, for IASD-fluxes, $N_{\rm flux}$ is negative. The largest
possible value for this solution is $N_{\rm flux}=-128$, which is achieved
whenever we choose one of the $b_i$ or $d_i$ equal to $\pm 8$ and the
others all equal to zero, or whenever we choose two coefficients to have the
absolute value of 8, but with differring signs, and all other coefficients
equal to zero.

\vskip0.2cm
\noindent
{\it (iv) $(3,0)$-flux}
\vskip0.2cm

This is an IASD-flux as well, again not consistent with the supergravity
equations of motion.

To obtain the $(3,0)$-part of the flux, we must set
$A_0=A_1=A_2=A_3=B_1=B_2=B_3=0$ or equivalently require that $\int G_3\wedge
\Omega=\int G_3\wedge \omega_{A1}=\int G_3\wedge \omega_{A2}=\int
G_3\wedge\omega_{A3}=\int G_3\wedge \omega_{B1}=\int G_3\wedge
\omega_{B2}=\int G_3\wedge\omega_{B3}=0$.
As a solution for $U^1=U^2=U^3=S=i$ we get
\eqn\eqdreinull{\eqalign{
{1\over{(2\pi)^2\alpha'}}G_{30}&=(d_0+id_3)\alpha_0+(d_3+id_0)\alpha_1\cr
&+(d_3+id_0)\alpha_2+(d_3+id_0)\alpha_3-(d_3+id_0)\beta^0\cr
&+(d_0-id_3)\beta^1+(d_0-id_3)\beta^2+(d_0-id_3)\beta^3,}}
which is the complex conjugate of our solution for the $(0,3)$-flux.
Expressed in the complex basis \cplxbase, the solution takes the form
\eqn\eqzdreinullcplx{\eqalign{
{1\over{(2\pi)^2\alpha'}}G_{30}&=(d_0-id_3)\om_{B0}.}}
In the case of $(3,0)$-flux, 
$$N_{\rm flux}=-4|B_0|^2=-4(d_0^2+d_3^2).$$ 
Here, the largest possible value for this solution is $N_{\rm flux}=-256$,
which we get for the same choices of coefficients as in the $(0,3)$-case.

\subsec{Examples of type $IIB$ orientifold models with non--vanishing fluxes}

In order to satisfy the supergravity equations of motion, the flux combination \fluxcomb\
has to obey the imaginary self--duality condition 
$\star_6 G=+i\ G$
\GKP.
This condition ensures the existence of a solution for the metric and
$4$--form. It also has crucial implications for the possible supersymmetry
breaking scenarios and for the choice of four--dimensional space--time.
From \fluxdec\ we immediately see, that in this case no $F$--term contributions to
the scalar potential $V$ are produced, \ie $V_F=V_{flux}=0$.
Only the $D$--term contribution $V_{top} \sim N_{flux}$ is non--vanishing. As already 
mentioned in section 2, this term enters $V_D$, given in \total.

Note, as discussed in section 2, the quantization conditions for the fluxes force the 
coefficients to be multiples of eight. This implies $|N_{flux}|\geq 64$.
For ISD--fluxes we have $N_{flux}>0$. 
According to \eqq \modcfour, consistent solutions, \ie vanishing Ramond tadpoles, 
may be possible without $D3$--branes, \ie $N_{D3}=0$. 
In that case, the $O3$--planes and/or non--trivial internal gauge bundles on the $D7$--brane 
world--volume may balance the positive $D3$--brane
charge introduced by the ISD--fluxes. This corresponds to the initial step, proposed in \KKLT,
to fix the dilaton and complex structure moduli only by ISD--fluxes. 
With that amount of flux, \ie  $N_{flux}=128$ and the stacks of $D7$--branes, displayed in Table 2,
the conditions \modcfour\ and \ceight\ are met with $N_{D3}=0$.
\vskip0.5cm
{\vbox{\ninepoint{
\def\ss#1{{\scriptstyle{#1}}}
$$
\vbox{\offinterlineskip\tabskip=0pt
\halign{\strut\vrule#
&~$#$~\hfil
&\vrule#
&~$#$~\hfil
&\vrule#&\vrule#
&~$#$~\hfil
&~$#$~\hfil
&~$#$~\hfil
&~$#$~\hfil
&~$#$~\hfil
&\vrule#
&~$#$~\hfil
&\vrule#
&~$#$~\hfil
&\vrule#  
\cr
\noalign{\hrule}
&
{\rm Stack}
&&
{\rm Gauge\ group}
&&&
(m^1,n^1)
&&
(m^2,n^2)
&&
(m^3,n^3)
&&
N_a
&
\cr
\noalign{\hrule}
\noalign{\hrule}
&
1,2,3,4
&&
SO(8)^4
&&&
(1,0)
&&
(-1,0)
&&
-
&&
4\times 4\ D7
&
\cr
&
5,6,7,8
&&
SO(8)^4
&&&
(1,0)
&&
-
&&
(-1,0)
&&
4\times 4\ D7
&
\cr
&
9,10,11,12
&&
SO(14)^4
&&&
-
&&
(1,0)
&&
(-1,0)
&&
4\times 7\ D7
&
\cr
&
13
&&
U(12)
&&&
-
&&
(1,2)
&&
(1,-2)
&&
12\ D7
&
\cr
\noalign{\hrule}}}$$
\vskip-10pt
\centerline{\noindent{\bf Table 2:}
{\sl $D7$--brane configuration allowing for  $N_{flux}=128$: wrapping numbers $m^j$,}}
\centerline{\sl  internal flux numbers $n^j$ and amount of
supersymmetry preserved.}}
\vskip10pt}}
\vskip-0.5cm \ \br
The first twelve  stacks (and their mirrors)
of $D7$--branes are located at the twelve $O7$--planes. Besides, the first eight stacks 
cancel tadpoles from the $C_8$--form locally. 
In order for the last stack to preserve the
supersymmetry condition \susy, we need to fix the K\"ahler moduli: $\Tc^2=\Tc^3$.
An example leading to $N_{flux}=128$ has been presented in 
subsection 2.3(i).\foot{Note
that N=1 supersymmetry is only mutually
preserved within the first three stacks of $D7$--branes. However,
the fourth stack, being itself N=2 supersymmetric, is
non--supersymmetric w.r.t. the other three. An alternative, completely
supersymmetric model with fluxes is provided by keeping the first
three stacks of $D7$--branes with $N_1=N_2=N_3=16$ and
$N_{flux}=32$. 
Hence, for this model fluxes through twisted cycles are needed
\doubref\BLT\CU}

As a final step, it has been suggested in \KKLT\ to include anti--$D3$ branes , \ie $N_{D3}<0$. 
This requires additional positive flux contributions in order for  \eqqs \modcfour\ and \ceight\
now to be satisfied. However, in that case, cancellation of $RR$--tadpoles does no longer
imply the absence of the $NSNS$--tadpoles, since anti--$D3$--branes contribute
to \modcfour\ with a negative sign in contrast to \PDthree. Hence, a positive 
$D$--term potential remains due to the uncancelled $NSNS$--tadpoles.
In fact, for anti--$D3$--branes with ISD--fluxes, soft--supersymmetry breaking terms
are generated and supersymmetry is broken \CIU. This may be a favorable scenario (\cf \KKLT),
since the soft--supersymmetry breaking terms fix the positions of the anti--$D3$--branes.

According to \fluxdec, only fluxes $G_3$ of IASD--type give rise to an $F$--term 
contribution $V_F$ to the scalar potential $V$ in $D=4$ space--time dimensions. 
The latter is positive semidefinite.
On the other hand, $N_{flux}<0$ for IASD--fluxes and generically, $D3$--branes are needed to
fulfill \modcfour. The supergravity equations of motions are inconsistent for such configurations
\GKP.
This fact is related to uncancelled $NSNS$--tadpoles and a positive scalar potential $V$.

ISD--fluxes do not lead to a potential for the $D3$--matter fields $C_i^3$ \doubref\CIU\JGL.
Hence in that case, those moduli remain undetermined and one has to find other mechanism to
stabilize them. Contrarily, $IASD$--fluxes do no fix the matter fields $C_i^3$ of 
anti--$D3$--branes.

\newsec{The effective action of toroidal \tb orbifold/orientifolds with $D3$--and $D7$--branes}

It is well known, that any N=1 supergravity action in four space--time dimensions
is encoded by three functions, namely the K\"ahler potential $K$,
the superpotential $W$, and the gauge kinetic function $f$    \CremmerEN.
When such an effective action arises from a higher dimensional string theory, 
these three functions usually depend (non--trivially) 
on moduli fields describing the background of the present string model.
In string compactifications with $D$--branes, one has to distinguish between closed string moduli
from the bulk, and open string moduli related to the $D$--branes.
The tree--level effective action describing the couplings of open and
closed string moduli up to second derivative order has been determined 
for toroidal orbifold/orientifold models with $D3/D7$--branes or $D5/D9$--branes 
and $2$--form fluxes by computing string scattering amplitudes \LMRS. 
The low--energy action for only the closed string bulk moduli has been discussed 
for \tb Calabi--Yau orientifolds in \doubref\JGL\GL.

There are several types of moduli fields in a type $IIB$ orientifold
compactification 
with $D$--branes. 
The closed string moduli fields arise from dimensional reduction of the 
bosonic part $(\phi,g_{MN},b_{MN},C_0,C_2,C_4)$ of the N=2 supergravity multiplet in $D=10$
after imposing the orientifold and orbifold action.
In the following, let us concentrate on \tb\ toroidal orbifolds/orientifolds 
with $D3$-- and $D7$--branes. Hence, the spectrum has to be invariant under both
the orientifold action $\Om(-1)^{F_L}I_6$ and the orbifold group $\Gamma$.
Before applying the orbifold twist $\Gamma$, the untwisted sector constitutes the states
invariant under $\Om(-1)^{F_L}I_6$: $\phi,g_{ij},b_{\mu i},C_0, C_{\mu i}, C_{ijkl},C_{\mu\nu ij},
C_{\mu\nu\rho\sigma}$. The fields $\phi\equiv \phi_{10}$ and $C_0$ constitute the 
universal dilaton field \dilaton.
The field $C_{\mu\nu\rho\sigma}$ gives rise to a tadpole to be cancelled by
the mechanism described in the previous section.
Imposing on all the states encountered above the orbifold 
action\foot{The orbifold group $\Gamma=\IZ_2\times\IZ_2$, with the generators given in \generator, 
leads to the following bosonic fields from the untwisted sector:
$\phi,C_0,C_{\mu\nu\rho\sigma},C_{\mu\nu j\ov j},C_{i\ov i j \ov j}$, 
and the internal metric $g$ has to have the block--diagonal structure $g=\otimes_{j=1}^3 g_j$,
with $g_j$ given in \eqq \metric. The latter gives rise to $9$ metric moduli, 
out of which the three complex structure moduli $\Uc^j$ \complexs\ 
and three real K\"ahler moduli $\im\Tc^j$ \kaehler\ are built. The latter are complexified
with the three axions $a^j=\int_{T^{2,k}\times T^{2,l}} C_{k\ov k l\ov l}\ , 
(j,k,l)=\overline{(1,2,3)}$. In $D=4$, these scalars 
are dual to the anti--symmetric $2$--tensors  $\int_{T^{2,j}} C_{\mu\nu j\ov j}$ and are eliminated
as a result of imposing self--duality on the self--dual $4$--form $C_4$. 
As already reported in the previous section,
the twisted sector contains $48$ additional complex structure moduli.} $\Gamma$ 
gives rise to the closed string untwisted sector. In addition there are twisted
moduli comprising the twisted $RR$--tensors.

Let us now come to the open string moduli fields. The massless untwisted moduli fields
originate from the $D=10$ gauge field $A_M$ reduced on the various $D$--branes.
The orientifold projection $\Om$ just determines the allowed  Chan--Paton gauge degrees of 
freedom at the open string endpoints.
For a stack of space--time filling $D3$--branes, we obtain $6$ real scalars 
$\phi^i\ ,\ i=4,\ldots, 9$ in the adjoint of the 
gauge group of the respective stack. These scalars describe the transversal 
movement of the $D3$--branes, \ie essentially the location of the $D3$--branes on the 
six--dimensional compactification manifold.
They may be combined into the three complex fields 
$C_i^3=\phi^{2i+2}+\Uc^i \phi^{2i+3}\ ,\ i=1,2,3$.
Furthermore, for a stack of $D7_3$--branes, which is wrapped around the $4$--cycle 
$T^{2,1}\times T^{2,2}$, we obtain the four Wilson lines $A_i\ ,\ i=4,5,6,7$ and 
two transversal coordinates $\phi^8,\phi^9$ in the adjoint representation. 
The latter describe the position of the $D7$--brane on the $2$--torus $T^{2,3}$.
Again, these six real fields may be combined into three complex fields $C^{7_3}_i\ ,\ i=1,2,3$.
After taking into account the other two $4$--cycles\foot{We assume the $4$--cycle to be a direct
product of two $2$--tori. At any rate, since $h_{(1,1)}=h_{(2,2)}=3$, these are the only 
$4$--cycles in the orbifold under consideration.}, 
on which other stacks of $D7$--branes may be wrapped, in total, 
we obtain the complex fields $C^{7_j}_i\ ,\ i,j=1,2,3$.
All the open  string fields described so far, give rise to complex scalars of 
the untwisted sector with at least N=2 supersymmetry.
A stack of  $D3$--branes gives rise to an N=4 super Yang--Mills theory on their world--volume,
provided the stack does not sit at an orbifold singularity.
Hence, together with their world--volume gauge fields, the scalars $C_i^3$ are organized 
in $N=4$ vector multiplets. The supersymmetry on the world--volume of a 
$D7$--brane, which is wrapped around a generic supersymmetric $4$--cycle of a
CY--space, is N=2. The scalar $C^{7_i}_i$ describes one (complex) scalar of a vectormultiplet.
However, on the torus $T^6$, the world--volume theory on the $D7$--branes
preserves N=4 supersymmetry. This becomes clear because one obtains in
addition the scalars $C^{7_j}_i\ ,\ i\neq j$, which represent four real
N=2 hypermultiplet scalars.
Hence together with the N=2 vectormultiplet, the latter constitute one N=4 vectormultiplet.

However, N=2 and chiral N=1 fields come from the twisted sector.
The twisted matter fields $C^{37_a}$ 
originate from open strings stretched between the $D3$-- and $D7$--branes from the $a$--th
stack. Without $2$--form fluxes
on the internal coordinates of the $D7$--brane these strings have $DN$--boundary conditions 
w.r.t. to those four coordinates. With non--vanishing $2$--form fluxes on the $D7$--brane
world--volume, the
$DN$--boundary conditions 
become mixed boundary conditions, with one open string end respecting the fluxes on the 
$D7$--branes. Generically, these fields respect N=2 supersymmetry. However, there are twisted N=1 
matter fields $C^{7_a7_b}$ arising from open strings 
stretched between two different stacks $a$ and $b$ of $D7$--branes.

Let us now move on to the low--energy effective action describing the dynamics
of the various  moduli fields encountered above.
The encountered complex scalars $S,T^j,U^j$ give rise to the closed string moduli space.
Since these fields live in the bulk, they constitute the bulk--moduli space. The latter
is a K\"ahler manifold, with the 
corresponding K\"ahler potential $\hat K$ given by \ABFPT:
\eqn\Khut{
\kappa_4^2\hat{K}=-\ln(S-\ov{S})-\sum_{i=1}^3\ln(T^i-\ov{T}^i)-
\sum_{i=1}^3\ln(U^i-\ov{U}^i).}
The untwisted open string moduli describe either the displacement 
transverse to the $D$--brane world--volume or the breaking of the gauge group by Wilson lines.
It is justified to expand the (full) K\"ahler potential $K$ and superpotential $W$
around this minimum $C_i=0$. In the case, that we have one stack of space--time filling
$D3$--branes and several stacks of $D7$--branes wrapped around different $4$--cycles 
(and being transverse to one torus $T^{2,j}$), the
K\"ahler potential takes the following expansion 
(for large K\"ahler moduli, which corresponds to the 
supergravity approximation under consideration):
\eqn\KK{\eqalign{
K(M,\ov M, C, \ov C)=\hat{K}(M, \ov M)&+
\sum_{i=1}^3 G_{C^3_i\ov C^3_i}(M,\ov M)\ C^3_i\ \ov C^3_i\cr
&+\sum_{a}\sum_{j=1}^3\sum_{i=1}^3 G_{C^{7_a,j}_i\ov C^{7_a,j}_i}(M,\ov M)\ C^{7_a,j}_i
\ov C^{7_a,j}_i\cr
&+\sum_{a} G_{C^{37_a}\ov C^{37_a}}(M,\ov M)\ C^{37_a}\ \ov C^{37_a}\cr
&+\sum_{a\neq b} G_{C^{7_a7_b}\ov C^{7_a7_b}}(M,\ov M)\ C^{7_a7_b}\ \ov C^{7_a7_b}+\Oc(C^4)\ .}}
Here, $M$ collectively accounts for the closed string moduli fields $S,\ T^i,\ U^i$, and
$C$ for the open string moduli, \ie matter fields.
The index $a$ denotes a particular stack of $D7$--branes, which has a fixed transverse
torus $T^{2,j}$.
Furthermore, the holomorphic superpotential $W$ takes the form:
\eqn\super{\eqalign{
W(M, C)&=\hat{W}(M)+C^3_1C^3_2C^3_3+\sum_a\sum_{j=1}^3 C^{7_a,j}_1 C_2^{7_a,j} C_3^{7_a,j}\cr
&+\sum_{I,J,K} Y_{IJK}(U^i)\ C_IC_JC_K+ \Oc(C^4)\ .}}
Here, the $C_I$ collectively account for all other combinations of matter
fields, in particular the twisted matter fields $C^{37_a},C^{7_a7_b}$.
The Yukawa couplings $Y_{IJK}(U^i)$, depending only on the complex
structure moduli $U^i$, have been determined in {\it Ref.} \YUK.
In vacua without fluxes, we have $\hat{W}(M)=0$ as a result of the flatness of the closed 
string moduli.
The closed string superpotential $\hat{W}(M)$, 
which is induced by the $3$--form flux has been already presented in \eqq \TVW.
Obviously, it is only non--vanishing for $(0,3)$--flux.

\subsec{Matter field metrics of untwisted open strings}

The metrics $G_{C^3_i\ov C^3_j}$ and $G_{C^{7,j}_i\ov C^{7_j}_i}$ for the untwisted matter 
fields $C^3_i$ and $C^{7_j}_i$ (of a particular $D7$--brane stack $a$) 
may be obtained from the following differential equation \LMRS:
\eqn\diff{
\partial_{\im\Tc^j}\ G_{C_i\ov C_i}={{D^j+\ov D^j}\over{4{\rm
Im}\Tc^j}}\ (1-2\delta^{ij})\ G_{C_i\ov C_i}\ .}
The latter is derived from a string scattering 
amplitude of two matter fields $C_i,\ C_k$ and the closed string modulus $\Tc^j$. 
There is no coupling between matter fields referring to different planes, \ie
$G_{C_i\ov C_j}=0\ ,\ i\neq j$ as a result of internal charge conservation \LMRS.
This fact justifies our ansatz for the expansion of K\"ahler potential \KK\ ex post facto.
The ``matrix'' $D$ depends on whether Dirichlet or Neumann boundary conditions are
imposed on the open string fields attached to the $D$--brane. In other words, the matrix
specifies, which of the matter fields $C_i^3, C_i^{7,j}$ we are considering 
in the scattering process.
For untwisted $D7$--brane matter fields $C_i^{7,j}$, the matrix $D$ also encodes 
the wrapping and flux properties of the $D7$--brane. 
The $D3$--brane case is particularly simple. In that case, all six internal open string
coordinates respect Dirichlet boundary conditions, \ie $D^i=-1$, and the differential
equation \diff\ yields:
\eqn\drei{
G_{C_i^3\ov C_i^3}\sim \ap^{-3/2}\ e^{-\phi_4}\ 
\ \fc{1}{\Uc^i-\ov \Uc^i}\ 
\sqrt{\fc{(\Tc^i-\ov \Tc^i)}{(\Tc^k-\ov \Tc^k)(\Tc^l-\ov \Tc^l)}}\ \ \ ,\ \ \ 
(i,k,l)=\overline{(1,2,3)}\ .}
Expressed in the field--theory moduli \fieldT\ and \fieldS, we obtain:
\eqn\Drei{
G_{C^3_i\ov C^3_i}=\fc{-\kappa_4^{-2}}{(U^i-\ov U^i)\ (T^i-\ov T^i)\ }\ \ \ ,\ \ \ i=1,2,3\ .}
The extra $U$--dependence comes from considering a four--point scattering
amplitude \LMRS.
Let us now move on to the untwisted $D7$--matter fields $C_i^{7,j}$.
For concreteness, let
us consider the fields $C^{7,3}_i$, \ie we shall discuss the case of a $D7$--brane, which 
is transversal to the third torus $T^{2,3}$. In this specific case, we find:
\eqn\Ds{\eqalign{
D^1&={{\im\Tc^1-if^1}\over{{\im\Tc^1+if^1}}}\ ,\cr
D^2&={{\im\Tc^2-if^2}\over{{\im\Tc^2+if^2}}}\ ,\cr
D^3&=-1\ .}}
After substituting this and integrating the differential equation, we obtain
the following metrics:
\eqn\metrics{\eqalign{
G_{C^{7,3}_1\ov C_1^{7,3}}&\sim e^{-\phi_4}\ \fc{1}{\Uc^1-\ov \Uc^1}\ 
\sqrt{{{{\Tc}^1-\ov\Tc^1}\over{(\Tc^2-\ov\Tc^2)(\Tc^3-\ov\Tc^3)}}}\ {{|\im\Tc^2+if^2|}
\over{|\im\Tc^1+if^1|}}\ ,\cr
G_{C^{7,3}_2\ov C^{7,3}_2}&\sim e^{-\phi_4}\ \fc{1}{\Uc^2-\ov \Uc^2}\ 
\sqrt{{{{\Tc}^2-\ov\Tc^2}\over{(\Tc^1-\ov\Tc^1)(\Tc^3-\ov\Tc^3)}}}\ {{|\im\Tc^1+if^1|}
\over{|\im\Tc^2+if^2|}}\ ,\cr
G_{C^{7,3}_3\ov C^{7,3}_3}&\sim e^{-\phi_4}\ \fc{1}{\Uc^3-\ov \Uc^3}\ 
\sqrt{{{{\Tc}^3-\ov\Tc^3}\over{(\Tc^1-\ov\Tc^1)(\Tc^2-\ov\Tc^2)}}}\ \lf|\im \Tc^1+if^1\ri|\ 
\lf|\im\Tc^2+if^2\ri|\ .}}
Expressed through the moduli in the field theory basis, the metric reads
\eqn\metricsfield{\eqalign{
G_{C^{7,3}_1\ov C^{7,3}_1}&={-\kappa_4^{-2}\over {(U^1-\ov U^1)\ (T^2-\ov{T}^2)}}\ 
{{|1+i\tilde{f}^2|}\over{|1+i\tilde{f}^1|}}\ ,\cr
G_{C^{7,3}_2\ov C^{7,3}_2}&={-\kappa_4^{-2}\over {(U^2-\ov U^2)\ (T^1-\ov{T}^1)}}\ 
{{|1+i\tilde{f}^1|}\over{|1+i\tilde{f}^2|}}\ ,\cr
G_{C^{7,3}_3\ov C^{7,3}_3}&={-\kappa_4^{-2}\over{(U^3-\ov U^3)\ (S-\ov{S})}}\ 
|1-\tilde{f}^1\tilde{f}^2|\ ,\cr
G_{C_i\ov C_k}&=0,\quad i\neq k\ ,}}
where $\tilde{f}^i={f^i\over{\rm Im} \Tc^i}$ is the physical 2--form flux. The other cases
$G_{C^{7,j}_i\ov C^{7,j}_i}$ with $j=1,2$ are obtained from the above results by permuting
fields. After putting our results together, the K\"ahler potential \KK\ for the untwisted closed
string sector becomes up to second order in the open string matter fields
\eqn\KW{\eqalign{
\kappa_4^2\ K(M,\ov M, C, \ov C)&=-\ln(S-\ov{S})-
\sum_{i=1}^3\ln(T^i-\ov{T}^i)-\sum_{i=1}^3\ln(U^i-\ov{U}^i)\cr
&+\sum_{i=1}^3 \ \fc{|C^3_i|^2}{(T^i-\ov T^i)(U^i-\ov U^i)}+
\sum_{a}\sum_{i=1}^3 d_{ikl}\ \fc{|C^{7_a,i}_i|^2}{(S-\ov S)(U^i-\ov U^i)}\ 
\lf|1-\tilde{f}^k\tilde{f}^l\ri|\ \cr
&+\sum_{a}\sum_{j=1}^3\sum_{i=1}^3 d_{ijk}\ \fc{|C^{7_a,j}_i|^2}{(T^k-\ov T^k)\ (U^i-\ov U^i)}\ 
\lf|\fc{1+i\tilde f^k}{1+i\tilde f^i}\ri|\ .}}
Here, we have introduced the tensor $d_{ijk}$ which is 1 for $(i,j,k)$ a 
permutation of $(1,2,3)$ and 0 otherwise.
There is one comment in order. As already anticipated, the $D3$--brane moduli $C_j^3\ ,\ 
j=1,2,3$ describe scalars of an N=4 vector multiplet and the $D7$--brane moduli $C_i^{7,i}$ 
are N=2 vectormultitplet scalars. 
The N=4 vector multiplet of the $D3$--brane may be split into 
one N=2 hypermultiplet in the adjoint and one N=2 vectormultiplet. The latter describes
the relative position of the $D3$--brane to that of the $D7$--brane in the space transversal to the
$D7$--brane. Hence, for a given internal index $i$ the fields $C_i^{7,i}$ and $C_i^3$ represent 
N=2 vectormultiplet scalars and we expect their metrics to be deducible from a common
N=2 prepotential $\Fc$. Indeed, for the case of vanishing $2$--form fluxes, 
from the N=2 trilinear prepotential $\Fc$ of special geometry \ADFT:
\eqn\prep{
\Fc(S,T^i,U^i,C_i^3,C_i^{7_a,i})= S\ [\ T^i U^i+\h (C_i^3)^2\ ]+
\h \sum_a T^i (C_i^{7_a,i})^2\ .}
we derive the untwisted sector K\"ahler potential 
\eqn\kaehlerf{\eqalign{
-\ln\lf[(S-\ov S)(T^i-\ov T^i)(U^i-\ov U^i)
+\h (S-\ov S)(C_i^3-\ov C_i^3)^2+\h\sum_a(T^i-\ov T^i) (C_i^{7_a,i}-\ov C_i^{7_a,i})^2\ri]}}
describing to the order of $C\ov C$ a part of the second line of \KW\ (in the case without
$2$--form fluxes).
In addition, from \eqq \kaehlerf\ there follows an $H$--term 
\eqn\Hdrei{
H_{ii}=-\fc{\kappa_4^{-2}}{(U^i-\ov U^i)(T^i-\ov T^i)}}
for the $D3$--brane moduli $C_i^3$. For the $D7$--brane scalars $C_i^{7,i}$ the  
following $H_{ii}$--term is generated:
\eqn\Hacht{
H_{ii}=-\fc{\kappa_4^{-2}}{(U^i-\ov U^i)(S-\ov S)}\ .}
However, in the more general case of non--vanishing $2$--form fluxes and several
stacks of $D7$--branes wrapping different $4$--cycles, the K\"ahler 
potential has to be given in the form \KW. There is no obvious disentangling of moduli
field according to the N=2 supersymmetry on a given $D7$--brane. Hence in that general case
we do not encounter $H$--terms, \ie $H_{ii}=0$.

\subsec{Matter field metrics for $1/4$ BPS brane configurations}

Let us now come to the K\"ahler metrics of the twisted sector matter fields $C^{7_a7_b}$. 
These fields arise from massless open strings stretched between two different stacks $a,b$ of 
$D7$--branes of $D7$--branes. 
These bosonic fields, arising from the open string $NS$--sector,  
build chiral multiplets, with their fermionic partners arising from the twisted $R$--ground
state.
Generically, the open strings respect only N=1 supersymmetry due to the different $2$--form flux
distributions on the $D7$--branes. In the dual \ta picture, these open strings are 
stretched between two intersecting $D6$--branes, with angles $\th^j_{ab}\ , \ j=1,2,3$. 
The matter field metrics for scalar matter fields arising from those
open strings has been calculated in {\it Ref.} \LMRS.
Essentially, we have to translate these results into the \tb picture with non--vanishing 
$2$--form fluxes.
For concreteness, let us consider the case of an open string ``stretched'' between two
$D9$--branes $D9_a$ and $D9_b$ with the $2$--form fluxes $f_a^j,f_b^j\ ,\ j=1,2,3$ on
their internal world--volume. In that case the matter field metric takes the following form:
\eqn\ninenine{
G_{C^{9_a9_b}\ov C^{9_a9_b}}=\kappa_4^{-2}\ \prod_{j=1}^3 (U^j-\ov U^j)^{-\th^j_{ab}}\ 
\sqrt{\fc{\Gamma(\th^j_{ab})}{\Gamma(1-\th^j_{ab})}}\ \ \ ,\ \ \ \th_{ab}^j\neq 0,1\ .}
Here the angle $\th_{ab}^j$, reminiscent from the \ta description, encodes 
the two flux components on the different stacks $a$ and $b$ of $D9$--branes:
\eqn\relativflux{
\th_{ab}^j=\fc{1}{\pi}\ \lf[\ \arctan\lf(\fc{f_b^j}{\im(\Tc^j)}\ri)-
\arctan\lf(\fc{f_a^j}{\im(\Tc^j)}\ri)\ \ri]\ .}
Generically, for two stacks of $D7$--branes $a$ and $b$, which wrap different $4$--cycles and
whose $2$--form fluxes fulfill \susy, there is always a 
non--vanishing relative ``flux'' $\th_{ab}^j$ in each plane, \ie $\th_{ab}^j\neq 0,1$.
In other words, these stacks are N=1 supersymmetric. Hence, in that case \eqq \ninenine\
directly applies and we obtain
\eqn\siebensieben{
G_{C^{7_a7_b}\ov C^{7_a7_b}}=\kappa_4^{-2}\ \prod_{j=1}^3 (U^j-\ov U^j)^{-\th_{ab}^j}\ 
\sqrt\fc{\Gamma(\th_{ab}^j)}{\Gamma(1-\th_{ab}^j)}\ ,}
with $\th_{ab}^j$ given in \relativflux.

However, if $\th^j_{ab}=0,1$ for one complex plane $j$, the two stacks $a$ and $b$ preserve
N=2 supersymmetry and \ninenine\ does not directly apply to these cases.
This case will be discussed in the next subsection.

\subsec{Matter field metrics for $1/2$ BPS brane configurations}

In this subsection, we shall derive the metrics for matter fields arising from open
strings stretched between two branes $a$ and $b$, which preserve N=2 space--time supersymmetry.
This case has not been discussed in \LMRS\ and a direct application of
the formula \ninenine, valid for the N=1 case, is not obvious\foot{Note, that a similar problem
arises in the case of one--loop gauge threshold corrections \LS. There, the one--loop correction
arising from open string N=1 sectors is not directly related to the
contribution stemming  from
the N=2 sectors.}.
Generically, these cases arise, if both branes $a$ and $b$ have vanishing fluxes $f^j=0$ 
in one and the same internal complex plane $j$.
In the dual \ta picture, this corresponds to the relative angle $\th_{ab}^j=0,1$ in that plane.
In particular, this is  true for the metrics $G_{C^{37_a}\ov C^{37_a}}$. 
In addition, two $D7$--branes $D7_a$ and $D7_b$ without
internal $2$--form fluxes, one transversal to the torus $T^{2,a}$ and the other transversal
to the torus $T^{2,b}$, are $1/2$ BPS.

In the following, we shall assume for concreteness, that both branes have their vanishing
$2$--form fluxes in the third complex plane, \ie $f_a^3=0=f_b^3$.
The two open string coordinates referring to this plane obey either pure Dirichlet
or Neumann boundary conditions at both string ends.
W.r.t. the other two planes, the branes $a$ and $b$ 
may carry the $2$--form fluxes $f_a^j,f_b^j\ ,\ j=1,2$ on their internal world--volume. 
The latter are assumed to fulfill the supersymmetry condition \susy:
\eqn\susycondition{
\fc{f_x^2}{\im(\Tc^2)}=-\fc{f_x^1}{\im(\Tc^1)}\ \ \ ,\ \ \ x=a,b\ .}
In the dual \ta picture, the branes $a$ and $b$ describe two intersecting $D6$--branes.
Their relative angles $\th_{ab}^j=\th_b^j-\th_a^j$ 
(with $\tan(\pi\th_x^j)=\fc{f_x^j}{\im(\Tc^j)}\ ,\ x=a,b$)
are given by \relativflux, for $j=1,2$. Furthermore, $\th_{ab}^3=0$.
In order, for the two branes to preserve N=2 supersymmetry, these relative angles have to
fulfill:
\eqn\twosusy{
\th^1_{ab}+\th_{ab}^2=0\ \mod\ 1\ .}

Massless open strings stretched between brane $a$ and $b$ give rise to N=2 matter fields
from the twisted sector. In the $(-1)$--ghost picture, their vertex operator is given by \BDL:
\eqn\mattervertex{
V_{C_\th}^{(-1)}(z,k)=\lambda\ 
e^{-\phi(z)}\ \prod_{j=1}^2 s_{\th^j}(z)\ \sigma_{-\th^j}(z)\ e^{ik_\nu X^\nu(z)}\ .}
The bosonic twist fields $\sigma_{-\th^j}$ have conformal dimension $\h \th^j(1-\th^j)$ 
and the spin fields have dimension $\h(1-\th^j)^2$. With this information and after imposing 
\twosusy, it is straightforward
to check, that the operator \mattervertex\ has conformal dimension one. 
To derive the moduli dependence of the matter field metric $G_{C^{ab}\ov C^{ab}}$, we proceed
similarly as in \LMRS, \ie we calculate the three--point amplitudes 
$\vev{V^{(-1)}_{C_{\th}} V^{(-1)}_{\ov C_{\th}} V^{(0,0)}_{\Uc^j}}$ and the four--point 
amplitudes $\vev{V^{(-1)}_{C_{\th}} V^{(-1)}_{\ov C_{\th}} V^{(0,0)}_{\Tc^i} 
V^{(0,0)}_{\ov \Tc^j}}$ in the dual \ta picture, \ie $C_\th \simeq C^{ab}$.
For the case $j=3$, the internal part of the modulus vertex $V^{(0,0)}_{\Uc^j}$ decouples
from the twist fields.  Essentially, the amplitude 
$\vev{V^{(-1)}_{C_{\th}} V^{(-1)}_{\ov C_{\th}} V^{(0,0)}_{\Uc^j}}$ leads to the same contractions
as in the case of two matter fields and one dilaton field, given in \LMRS.
To this end, we find the differential equation (already translated into the \tb picture): 
\eqn\important{
\p_{\Tc_2^3} G_{C^{ab}\ov C^{ab}}=\fc{\pm i}{\Tc^3-\ov \Tc^3}\ .}
The two signs depend on whether Neumann or Dirichlet  boundary conditions are imposed on
the open string coordinates from  the third plane, respectively.
Hence \important\ gives rise to the following $\Tc^3$--dependence:
\eqn\giverise{
G_{C^{ab}\ov C^{ab}}\sim (\Tc^3-\ov \Tc^3)^{\pm \h}\ .}
On the other hand, for $j=1,2$, the amplitude 
$\vev{V^{(-1)}_{C_{\th}} V^{(-1)}_{\ov C_{\th}} V^{(0,0)}_{\Uc^j}}$ yields the same differential
equation as found in \LMRS\ for each plane separately. 
Hence the total dependence on the K\"ahler moduli becomes
\eqn\totalkaehler{
G_{C^{ab}\ov C^{ab}}\sim (\Tc^3-\ov \Tc^3)^{\pm \h}\ \prod_{j=1}^2 
\sqrt{\fc{\Gamma(\th_{ab}^j)}{\Gamma(1-\th_{ab}^j)}}\ ,}
with the angles given in \eqq \relativflux.
To determine the complex structure dependence of the metric $G_{C^{ab}\ov
C^{ab}}$,  we
calculate the amplitude $\vev{V^{(-1)}_{C_{\th}} V^{(-1)}_{\ov C_{\th}} V^{(0,0)}_{\Tc^i} 
V^{(0,0)}_{\ov \Tc^j}}$ in the dual \ta picture. 
Clearly, the amplitude is vanishing for $i\neq j$ due to internal charge conservation.
Furthermore, for $i=1,2$, \ie the $\Tc^i$--moduli are from those two planes, for which
$\th_{ab}^i\neq 0$, our calculation boils down to the case discussed in \LMRS\ and yields
\eqn\find{
\vev{V^{(-1)}_{C_{\th}} V^{(-1)}_{\ov C_{\th}} V^{(0,0)}_{\Tc^i} 
V^{(0,0)}_{\ov \Tc^i}}\sim \fc{st}{u}+s\ \th^i+\ldots\ \ \ ,\ \ \ i=1,2\ ,}
with the three kinematic invariants $s,t$ and $u$. The dots stand for
higher orders in the space--time momentum, which are not relevant to us.
On the other hand, for $i=3$, we find:
\eqn\findi{
\vev{V^{(-1)}_{C_{\th}} V^{(-1)}_{\ov C_{\th}} V^{(0,0)}_{\Tc^i} 
V^{(0,0)}_{\ov \Tc^i}}\sim \fc{st}{u}+\ldots\ \ \ ,\ \ \ i=3\ .}
To the order in the momentum, which is displayed on the r.h.s. of both equations
\find\ and \findi, both sides have to reproduce the $\sigma$--model result
$G_{C^{ab}\ov C^{ab}}\ G_{\Tc^i\ov \Tc^i} \fc{st}{u}+s\ R_{C^{ab}\ov C^{ab}\Tc^i\ov \Tc^i}$.
This information allows us to completely fix the $\Tc^i$--dependence of the matter metric
$G_{C^{ab}\ov C^{ab}}$ in the $T$--dual \ta picture.

Finally, after putting all results together, we obtain 
the metric of massless matter fields originating from a $1/2$ BPS system of branes in \tb:
\eqn\finalmatter{
G_{C^{ab}\ov C^{ab}}=\ap^{-(1\pm \h)}\ e^{-\phi_4}\ (\Tc^3-\ov \Tc^3)^{\pm \h}\ 
\prod_{j=1}^2 (\Uc^j-\ov \Uc^j)^{-\th^j}\ \sqrt\fc{\Gamma(\th^j)}{\Gamma(1-\th^j)}\ .}

From this expression, we can deduce various special cases\foot{
Although not relevant to us in this article, for completeness, let us also discuss the special 
cases arising in \tb orientifolds with $D9$-- and $D5$--branes. 
First, we consider a $D5$--brane, wrapped around the third torus $T^{2,j}$ and a 
$D9$--brane wrapped around the full six--torus $T^6$. This system, which is $1/2$ BPS, 
preserves N=2 space--time supersymmetry. W.r.t. the third torus, open strings
have Neumann boundary conditions. On the other hand, in the dual \ta picture, 
the two branes intersect at the angles $\pi/2$ within the other two internal planes, 
\ie $\th_{ab}^j=\h\ ,\ j=1,2$. After recalling the relation between the moduli 
fields $\Tc^j$ in the string basis vs. field--theory basis, namely 
$\im(S)=(2\pi)^{-1}\ap^{-3/2}e^{-\phi_4}\ \sqrt{\Tc_2^1\Tc_2^2\Tc_2^3}$ and
$\im(T^j)=(2\pi)^{-1}\ap^{1/2}e^{-\phi_4}\ \sqrt{\fc{\Tc_2^j}{\Tc_2^k\Tc_2^l}}\ ,\ 
(j,k,l)=\overline{(1,2,3)}$ for \tb orientifolds with $D9$ and $D5$--branes, we
obtain (with $\kappa_4^{-2}=\fc{e^{-2\phi_4}}{\pi\ap}$):
\eqn\ninefive{\eqalign{
G_{C^{95_3}\ov C^{95_3}}&=2^{-1/2}\ i^{-3/2}\ 
\ap^{-3/2}\ \ e^{-\phi_4}\ (\Tc^3-\ov \Tc^3)^{1/2}\ 
\fc{1}{(\Uc^1-\ov \Uc^1)^{1/2}\ (\Uc^2-\ov \Uc^2)^{1/2}}\cr 
&=\fc{\kappa_4^{-2}}{(T^1-\ov T^1)^{1/2}(T^2-\ov T^2)^{1/2}}\ 
\fc{1}{(U^1-\ov U^1)^{1/2}\ (U^2-\ov U^2)^{1/2}}\ .}}
Furthermore, for two $D5$--branes, with one wrapping the torus $T^{2,1}$ and the other
one wrapping the torus $T^{2,2}$ we have pure Dirichlet boundary conditions w.r.t
the open string coordinates from the third plane. Again, in the dual \ta picture, 
the two branes intersect at the angles $\pi/2$ within the other two internal planes, 
\ie $\th_{ab}^j=\h\ ,\ j=1,2$. Hence, from \eqq \finalmatter\ we deduce:
\eqn\fivefive{\eqalign{
G_{C^{5_15_2}\ov C^{5_15_2}}&=2^{1/2}\ i^{-1/2}\ 
\ap^{-1/2}\ \ e^{-\phi_4}\ (\Tc^3-\ov \Tc^3)^{-1/2}\ 
\fc{1}{(\Uc^1-\ov \Uc^1)^{1/2}\ (\Uc^2-\ov \Uc^2)^{1/2}}\cr 
&=\fc{\kappa_4^{-2}}{(S-\ov S)^{1/2}(T^3-\ov T^3)^{1/2}}\ 
\fc{1}{(U^1-\ov U^1)^{1/2}\ (U^2-\ov U^2)^{1/2}}\ .}}}, which are not captured by 
the formula \ninenine. The latter is valid for $1/4$ BPS systems.
For a space--time filling $D3$--brane and a $D7$--brane, transversal to the 
third torus, we get\foot{We use the translation rules \fieldT\ to go from the string--basis
to the field--theory basis.}:
\eqn\threesevenn{\eqalign{
G_{C^{37_3}\ov C^{37_3}}&=2^{1/2}\ i^{-1/2}\ \ap^{-1/2}\ \ e^{-\phi_4}\ (\Tc^3-\ov \Tc^3)^{-1/2}\ 
\fc{1}{(\Uc^1-\ov \Uc^1)^{1/2}\ (\Uc^2-\ov \Uc^2)^{1/2}}\cr 
&=\fc{\kappa_4^{-2}}{(T^1-\ov T^1)^{1/2}(T^2-\ov T^2)^{1/2}}\ 
\fc{1}{(U^1-\ov U^1)^{1/2}\ (U^2-\ov U^2)^{1/2}}\ .}}
In the case, that on the internal $D7$--brane world volume non--trivial $2$--from fluxes 
$f^1,f^2$ satisfying \susycondition\ are turned on, we obtain from \eqq \finalmatter:
\eqn\threesevennn{\eqalign{
G_{C^{37_3}\ov C^{37_3}}&=\fc{\kappa_4^{-2}}{(T^1-\ov T^1)^{1/2}(T^2-\ov T^2)^{1/2}}\ 
\prod_{j=1}^2 (U^j-\ov U^j)^{-\th^j}\ \sqrt{\fc{\Gamma(\th^j)}{\Gamma(1-\th^j)}}\cr
&=\fc{\kappa_4^{-2}}{(T^1-\ov T^1)^{1/2}(T^2-\ov T^2)^{1/2}}\ 
\fc{1}{(U^1-\ov U^1)^{\th^1}\ (U^2-\ov U^2)^{\th^2}}}}
The last equation follows from the N=2 supersymmetry condition \twosusy, \ie
$\th_{ab}^1=1-\th_{ab}^2\ \mod\ 1.$
Finally, the special case of two $D7$--branes $D7_1$ and $D7_2$ without
internal $2$--form fluxes, one transversal to the torus $T^{2,1}$ and the other transversal
to the torus $T^{2,2}$, are $1/2$ BPS. This corresponds to the case $\th_{ab}^j=1/2\ ,\ j=1,2$
and $\th_{ab}^3=0$, already discussed in \eqq \fivefive\ for a system of two $D5$--branes.
The metric of the corresponding open string matter field becomes:
\eqn\fivefiveee{\eqalign{
G_{C^{7_17_2}\ov C^{7_17_2}}&=2^{-1/2}\ i^{-3/2}\ 
\ap^{-1/2}\ \ e^{-\phi_4}\ (\Tc^3-\ov \Tc^3)^{1/2}\ 
\fc{1}{(\Uc^1-\ov \Uc^1)^{1/2}\ (\Uc^2-\ov \Uc^2)^{1/2}}\cr 
&=\fc{\kappa_4^{-2}}{(S-\ov S)^{1/2}(T^3-\ov T^3)^{1/2}}\ 
\fc{1}{(U^1-\ov U^1)^{1/2}\ (U^2-\ov U^2)^{1/2}}\ .}}

To conclude, in this subsection we have derived the metric for massless
matter fields originating from open strings stretched between
a space--time filling $D3$--brane and a $D7$--brane, transversal to the first torus $T^{2,1}$
with the $2$--form fluxes $f^2,f^3$ on its internal world--volume:
\eqn\dreisieben{
G_{C^{37_1}\ov C^{37_1}}=\fc{\kappa_4^{-2}}{(T^2-\ov T^2)^\h\ (T^3-\ov T^3)^\h}\ 
\fc{1}{(U^2-\ov U^2)^{\th^2}\ (U^3-\ov U^3)^{\th^3}\ }}
for the matter field metric referring to an  open string stretched between a 
$D3$--brane and a $D7$--brane. The latter is located transversal to
the first complex plane and carries the $2$--form fluxes $f^2,f^3$ on its (internal) 
world--volume. The fluxes, which fulfill \susy, are encoded in the angles $\th^j$:
\eqn\fluxangle{
\th^j=\fc{1}{\pi} \arctan\lf(\fc{f^j}{\im(\Tc^j)}\ri)\ \ \ ,\ \ \ j=2,3\ .} 

\subsec{Gauge kinetic function}

Finally, the holomorphic gauge kinetic function, which encodes the gauge coupling
of a $D7$--brane, wrapped around the $4$--cycle $T^{2,k}\times T^{2,l}$ [with wrapping
numbers $m^k,m^l$ and the supersymmetric $2$--form fluxes $f^k,f^l$ (\cf condition \susy)] is:
\eqn\gauge{
f_{D7_j}(S,T^j)=|m^k m^l|\lf(T^j-\ap^{-2} f^kf^lS \ri) \ \ \ , \ \ \ (j,k,l)=\overline{(1,2,3)}\ .}
Furthermore, the gauge sector of a space--time filling $D3$--brane is encoded in the 
holomorphic function:
\eqn\gaugethree{
f_{D3}(S)=S\ .}
Note, that the gauge couplings of the $D3$--brane and
the various $D7$--branes are derived from the common holomorphic N=2 prepotential
\eqn\Prep{\eqalign{
\Fc(S,T^i,U^i,C_i^3,C_i^{7_a,i})&= S\ [\sum_{i=1}^3 T^i U^i+\h (C_i^3)^2]\cr
&+\h \sum_a  \sum_{(j,k,l)\atop=\overline{(1,2,3)}} (T^j-\ap^{-2} f^k f^l S) 
\ (C_j^{7_a,j})^2}}
as second derivative w.r.t. to the N=2 untwisted moduli fields $C_j^3,\ C_j^{7_a,j}$ 
describing vector multiplet scalars, 
\ie $g_{D7_a,j}^{-2}=\im \fc{\p^2 \Fc}{\p (C_j^{7_a,j})^2}$ and
$g_{D3}^{-2}=\im \fc{\p^2 \Fc}{\p (C_j^3)^2}$.
This represents 
a generalization of the expression \prep\ including all three complex planes $i=1,2,3$. 
This indicates, that the gauge couplings still obey N=2 supersymmetry, even though the whole
$D3/D7$--brane configuration is only N=1 supersymmetric.
In particular, even in the case of vanishing  $2$--form fluxes, the kinetic energy 
terms of the various bulk and brane moduli fields presented before, cannot be
derived  from \Prep.

\newsec{The effective action with 3-form fluxes turned on}

\subsec{Superpotential, $F$--terms and scalar potential}
We now want to obtain the low energy action with fluxes turned on. First, we
will look at those quantities that can be derived in the bulk.
The $F$-terms can be calculated from the K\"ahler potential \Khut\  and the
superpotential\ \TVW, they only feel the
bulk:
\eqn\Fterm{\ov{F}^{\ov{I}} = e^{\kappa_4^2\hat{K}/2}\ \hat{K}^{\ov{I}J} \ 
(\partial_J\hat{W}+\kappa_4^2\ \hat{W}\ \partial_J\hat{K})\ ,}
where the $I, \ J$ are taken to run over the dilaton $S$, the complex structure
moduli $U^i$  and the K\"ahler moduli $T^i$. With this, we can now calculate
the scalar potential:
\eqn\Vhut{\hat{V}=\hat{K}_{I\ov{J}}\ F^I\ov{F}^{\ov{J}}-3\ e^{\kappa_4^2\hat{K}}\ 
\kappa_4^2\ |\hat{W}|^2\ .}
With what we learned in section 2, it is now easy to write down the
superpotential explicitly for our case of $T^2\times T^2\times T^2$:
\eqn\WhT{\eqalign{
{1\over \lambda}\hat{W}&=(a^0-Sc^0)U^1U^2U^3-\{(a^1-Sc^1)U^2U^3+
(a^2-Sc^2)U^1U^3+(a^3-Sc^3)U^1U^2\}\cr
&-\sum_{i=1}^3(b_i-Sd_i)U^i-(b_0-Sd_0).}}
The explicit expressions for the $F$-terms are the following:
\eqn\ftermsT{\eqalign{
\ov{F}^{\ov{S}}&=(S-\ov{S})^{1/2}\prod_{i=1}^3(T^i-\ov{T}^i)^{-1/2}
\prod_{i=1}^3(U^i-\ov{U}^i)^{-1/2}\kappa_4^2{\lambda\over{(2\pi)^2\alpha'}}
\int\ov{G}_3\wedge \Omega\cr
&=\lambda\kappa_4^2(S-\ov{S})^{1/2}\prod_{i=1}^3(T^i-\ov{T}^i)^{-1/2}
\prod_{i=1}^3(U^i-\ov{U}^i)^{-1/2}\times\{(a^0-\ov{S}c^0)U^1U^2U^3-\cr
&-[(a^1-\ov{S}c^1)U^2U^3+(a^2-\ov{S}c^2)U^1U^3+(a^3-\ov{S}c^3)U^1U^2]\cr
&-\sum_{i=1}^3(b_i-\ov{S}d_i)U^i-(b_0-\ov{S}d_0)\}\ ,\cr
\ov{F}^{\ov{T}^i}&=(S-\ov{S})^{-1/2}(T^i-\ov{T}^i)^{1/2}(T^j-\ov{T}^j)^{-1/2}
(T^k-\ov{T}^k)^{-1/2}\prod_{i=1}^3(U^i-\ov{U}^i)^{-1/2}\kappa_4^2\
\hat{W}\cr
&=\lambda\kappa_4^2(S-\ov{S})^{-1/2}(T^i-\ov{T}^i)^{1/2}(T^j-\ov{T}^j)^{-1/2}
(T^k-\ov{T}^k)^{-1/2}\prod_{i=1}^3(U^i-\ov{U}^i)^{-1/2}\times\cr
&\times\{(a^0-Sc^0)U^1U^2U^3-[(a^1-Sc^1)U^2U^3+(a^2-Sc^2)U^1U^3+(a^3-Sc^3)U^1U^2]\cr
&-\sum_{i=1}^3(b_i-Sd_i)U^i-(b_0-Sd_0)\}\, ,\cr
\ov{F}^{\ov{U}^i}
&=(S-\ov{S})^{-1/2}(U^i-\ov{U}^i)^{1/2}(U^j-\ov{U}^j)^{-1/2}(U^k-\ov{U}^k)^{-1/2}
\prod_{l=1}^3(T^l-\ov{T}^l)^{-1/2}\times\cr
&\times\ \kappa_4^2{\lambda\over{(2\pi)^2\alpha'}} \int G_3\wedge
\omega_{A_i}\ ,\ \ i\neq j\neq k\ ,\ {\rm e.g.:}\cr
\ov{F}^{\ov{U}^1}&=\lambda\kappa_4^2(S-\ov{S})^{-1/2}(U^i-\ov{U}^i)^{1/2}
(U^j-\ov{U}^j)^{-1/2}(U^k-\ov{U}^k)^{-1/2}\prod_{l=1}^3(T^l-\ov{T}^l)^{-1/2}\times\cr
&\times\{((a^0-Sc^0)\ov{U}^1U^2U^3-[(a^1-Sc^1)U^2U^3+(a^2-Sc^2)\ov{U}^1U^3+(a^3-Sc^3)
\ov{U}^1U^2]\cr
&-[(b_1-Sd_1)\ov{U}^1+(b_2-Sd_2)U^2+(b_3-Sd_3)U^3]-(b_0-Sd_0)\}\ .}}
By looking at the $F$-terms, we see immediately that we only have a non-zero
$F^S$ if $G_3$ has a $(3,0)$-component. For $F^{T^i}$ to be non-zero, $G_3$
has to have a non-zero $(0,3)$-component. For the $F^{U^i}$ to be non-zero,
$G_3$ must have a $(1,2)$-component.

Now we are able to compute the expression for the scalar potential \Vhut. The part
coming from the $T$-moduli cancels with $-3e^{\hat{K}\kappa_4^2}\kappa_4^2|\hat{W}|^2$, so 
we are left with
\eqn\VhutT{\eqalign{\hat{V}&=\partial_S\partial_{\ov{S}}\hat{K}F^S\ov{F}^{\ov{S}}+
\sum_{i=1}^3\partial_{U_i}\partial_{\ov{U_i}}\hat{K}F^{U_i}\ov{F}^{\ov{U_i}}\cr
&={\lambda^2\kappa_4^2\over (2\pi)^4\alpha'^2}\ \lf(|S-\ov{S}|\prod_{i=1}^3|T^i-\ov{T}^i|
\prod_{i=1}^3|U^i-\ov{U}^i|\ri)^{-1}\left(|\int\ov{G}_3\wedge
\Omega|^2+\sum_{i=1}^3|\int G_3\wedge \omega_{A_i}|^2\right).}}
We can see immediately that $\hat{V}$ is zero unless $G_3$ has a $(3,0)$- or a
$(1,2)$-part, i.e. is IASD. We also see immediately, that for IASD-fluxes,
$\hat V$ is strictly positive.
When we express this through the complex coefficients, this formula looks even
nicer:
\eqn\VhutTc{\hat{V}=\lambda^2\kappa_4^2\ {\prod\limits_{i=1}^3|U^i-\ov{U}^i|\over|S-\ov{S}
|\prod\limits_{i=1}^3|T^i-\ov{T}^i|}\ \sum_{j=0}^3|B^j|^2.}
This ties in neatly with our observation from section 2: Let us examine
eq. \CPeven\ : Expressed with our complex coefficients, we find
\eqn\cpevenc{\int G_3\wedge \star_6\ov G_3\ \propto\  2\sum_{i=0}^3|B^i|^2+(\sum_{i=0}^3|A^i|^2-
\sum_{i=0}^3|B^i|^2)\ ,}
where the second term is obviously proportional to $N_{\rm flux}$, whereas
the first part corresponds to $V_{\rm flux}$, which is the contribution to the
scalar potential coming from the $F$-terms, which is exactly, what we have
calculated above.

The gravitino mass is given by
\eqn\gravitino{\eqalign{m_{3/2}&=e^{\kappa_4^2\hat{K}/2}\kappa_4^2\
\hat{W}=
\lambda\kappa_4^2\ (S-\ov{S})^{-1/2}\prod_{i=1}^3(T^i-\ov{T}^i)^{-1/2}
\prod_{i=1}^3(U^i-\ov{U}^i)^{-1/2}\times\cr
&\times\{(a^0-Sc^0)U^1U^2U^3-[(a^1-Sc^1)\ U^2U^3+(a^2-Sc^2)\ U^1U^3+(a^3-Sc^3)\ U^1U^2]\cr
&-\sum_{i=1}^3(b_i-Sd_i)\ U^i-(b_0-Sd_0)\}\ .}}

\subsec{Soft SUSY breaking terms}
The effective low energy supergravity potential in the standard
limit with $M_{\rm Pl}\to \infty$
with $m_{3/2}$ fixed is for $N=1$ supersymmetry \doubref\SOFTref\BIM:
\eqn\Veff{
V^{{\rm eff}} = {1\over 2} D^2
+  G^{C_I\ov C_I}\, |\partial_I W^{({\rm eff})}|^2+\ m^2_{I\ov{I},{\rm
soft}}\, C_{I}\ov C_I+{1\over 3}A_{IJK}C_IC_JC_K + {\rm h.c.}\ ,
}
with
\eqn\susyt{\eqalign{
D &= - g_I\kappa_4^2\ G_{C_I\ov C_I}C_I\ov C_I ,\cr
W^{({\rm eff})} &={1\over 3}\ e^{\kappa_4^2\ \hat{K}/2}  Y_{IJK}\,C_IC_JC_K.
}}
The $C_I$ are taken to run over the $C_i^3,\  C_i^{7a,j},\ C^{37,a},\ C^{7a7b}$,
where $i=1,2,3$, $a$ and $b$ run over the stacks of branes, and $j$ runs
over the  torus not being wrapped.
Furthermore, $g_I$ is the gauge coupling, which is related to the gauge kinetic function
$f_I(M)$ by $g_I^{-2}={\rm Im}(f_I(M))$. The respective gauge kinetic
functions are given in \gauge, \gaugethree. 

The diagonal structure of our metrics already results in some simplifications,
for example the purely diagonal structure of the scalar mass matrix. The fact
that we have $H_{ij}=0$ results in even more drastic simplifications: In our case, no $B$-term
$B_{IJ}C_IC_J$ 
appears in the effective scalar potential, and also no $\mu$-term $\half
\mu_{IJ}C_IC_J$ is generated
in $W^{({\rm eff})}$.

The soft supersymmetry breaking terms are
\eqn\susybr{\eqalign{
m^2_{I \ov{I},{\rm soft}}&= \kappa_4^2\ [\ (|m_{3/2}|^2 +\kappa_4^2\ \hat{V})\,G_{C_I\ov C_I} -
F^\rho \ov F^{\ov \sigma } R_{\rho {\ov \sigma} I \ov{I}}\ ]  ,\cr
A_{IJK} &= F^\rho D_\rho\ (e^{\kappa_4^2\hat{K}/2} Y_{IJK}),}}
where the Greek indices are running over $S,\ T^i,\ U^i$ and
\eqn\Ri{\eqalign{
R_{\rho \ov \sigma I \ov{I}} &= {{\partial^4 K}\over{\partial
C_I\partial\ov C_I\partial M_\rho\partial\ov M_\sigma}}-{{\partial^3 K}\over{\partial
C_I\partial
M_\rho\partial\ov C_K}}G^{\ov C^KC^K}{{\partial^3K}\over{\partial\ov C_I\partial
\ov M_\sigma\partial C_K}}\ ,\cr
D_\rho(e^{\kappa_4^2\ \hat{K}/2} Y_{IJK}) &=
\partial_\rho(e^{\kappa_4^2\hat{K}/2}  Y_{IJK}) + {1\over 2}
\kappa_4^2\hat{K}_\rho\ (e^{\kappa_4^2\hat{K}/2}
Y_{IJK})\cr
&-e^{\kappa_4^2\hat{K}/2}G^{\ov C_IC_I}\partial_\rho G_{C_I (\ov C_I}Y_{JK)I}.
 }}
The gaugino mass is
\eqn\gaugino{m_{gI}=F^\rho\ \partial_\rho\log({\rm Im}f_I),}
$f_I(M)$ being the gauge kinetic function.

\vskip0.2cm
\noindent
{\it Explicit calculation of the terms:}
\vskip0.2cm

We first look at $W^{({\rm eff})}$. We find
\eqn\weff{W^{({\rm eff})}={1\over
3}[(S-\ov{S})\prod_{i=1}^3(T^i-\ov{T}^i)\prod_{i=1}^3(U^i-\ov{U}^i)]^{-1/2}
\ Y_{IJK}\ C_IC_JC_K.}
 From eq. \super,  we know that $Y_{IJK}=\epsilon_{IJK}$ in the case of the
untwisted matter fields $C_i^3,C_i^{7,a}$ and the combination $\sum_a
C^{7_{a,1}7_{a,2}}C^{7_{a,2}7_{a,3}}C^{7_{a,3}7_{a,1}}$.

Before we can calculate the expressions for the scalar masses $m_{I\ov{I},{\rm soft}}$, we must
first find the explicit expressions for the curvature tensor. This is done in
Appendix \appB. We get
\eqn\msoft{\eqalign{
(m^{33}_{i\ov i})^2&=\kappa_4^2\ \lf[(|m_{3/2}|^2+\kappa_4^2\hat{V})\ G_{C^3_i\ov
C^3_i}-|F^{U^i}|^2R^{3}_{U^i\ov U^ii\ov{i}}-|F^{T^i}|^2R^{3}_{T^i\ov T^ii\ov{i}}\ri]\ ,\cr
(m^{7,j}_{i\ov i})^2&=\kappa_4^2\ \lf[(|m_{3/2}|^2+\kappa_4^2\hat{V})\ G_{C^{7,j}_i\ov
C^{7,j}_i}-\sum_{M,N}F^M\ov F^{\ov N}R^{7,j}_{M\ov N i\ov i}\ri]\ ,\cr
(m^{37_a})^2&=\kappa_4^2\ \lf[(|m_{3/2}|^2+\kappa_4^2\hat{V})\ G_{C^{37_a}\ov
C^{37_a}}-\sum_{M,N}F^M\ov F^{\ov N}R^{37_a}_{M\ov N}\ri]\ ,\cr
(m^{7a7b})^2&=\kappa_4^2\ \lf[(|m_{3/2}|^2+\kappa_4^2\hat{V})\ G_{C^{7a7b}\ov
C^{7a7b}}-\sum_{M,N}F^M\ov F^{\ov N}R^{7a7b}_{M\ov N}\ri]\ ,}}
where $M,\ N$ run over $S,\ T^i,\ U^i$.
The trilinear coupling is
\eqn\Aijk{\eqalign{
A_{IJK}&=i\prod_M(M-\ov M)^{-1}{\lambda\kappa_4^2\over
(2\pi)^2\alpha'}\left\{Y_{IJK}\int G_3\wedge \ov\Om+3\ Y_{IJK}\int \ov G_3\wedge \ov
\Om\right.\cr
&+\left.\sum_i\int \ov G_3\wedge \ov\om_{A_i}\ [Y_{IJK}-(U^i-\ov U^i)\ \p_{U^i}Y_{IJK}]\right\}\cr
&-i\prod_M(M-\ov M)^{-1/2}\ F^\rho\ G^{\ov C_IC_I}\ \partial_\rho
G_{C_I(\ov C_I}Y_{JK)I}\ .}}
The term $-\sum_i\int \ov G_3\wedge \ov\om_{A_i}(U^i-\ov U^i)\p_{U^i}Y_{IJK}$
appears, because general $Y_{IJK}$ may depend on the complex structure moduli.

Note the case where the $I,\  J,\  K$ refer to the 3-brane matter fields $C^3_i$: Then
the last term cancels the terms $3\int \ov G_3\wedge \ov
\Om$ and $\sum_i\int\ov G_3\wedge \ov\om_{A_i}$, and we are left with
$$A_{IJK}=i\, \epsilon_{IJK}\prod_M(M-\ov M)^{-1}{\lambda\kappa_4^2\over
(2\pi)^2\alpha'}\int G_3\wedge \ov\Om,$$
i.e. we only get a trilinear coupling from the $(3,0)$-flux, which agrees with
the results of \doubref\CIU\JGL. This is not true
for the other matter fields, as their metrics have a more complicated
dependence on the moduli.

The gauge couplings have been given in \eqq \gauge\ and \gaugethree. Through them, we obtain 
the gaugino masses:
\eqn\gauginoexpl{\eqalign{
m_{g, D7j}&=F^S\ {{-\alpha'^{-2}f^kf^l}\over{(T^j-\ov{T}^j)-\alpha'^{-2}f^kf^l(S-\ov{S})}}+
F^{T^j}\ {{1}\over{(T^j-\ov{T}^j)-\alpha'^{-2}f^kf^l(S-\ov{S})}},\cr
m_{g, D3}&=F^S{1\over(S-\ov S)}=-i\prod_M(M-\ov
M)^{-1/2}{\lambda\kappa_4^2\over (2\pi)^2\alpha'}\int G_3\wedge \ov\Om,
}}
with $k\neq l\neq j,\ $ $j$ being the torus not wrapped by the 7-brane.

We shall now examine the soft terms for the different flux components
turned on separately.
When examining the effective potential, we find that the $D$-term and
$W^{({\rm eff})}$ do not depend on the 3-form fluxes. 

\vskip0.2cm
\noindent
{\it (i) $(2,1)-$flux}
\vskip0.2cm

The $(2,1)$-flux fulfills the ISD condition $\star_6\,G_3=i\,G_3$
w.r.t. the Hodge operation in the internal (compact) six--dimensional compactification 
manifold $X_6$. 
For this flux component, the superpotential, 
the scalar potential in the bulk and all the $F$-terms are zero, as well as the gravitino mass 
$m_{3/2}$, the
soft terms and the gaugino mass $m_g$. This is no surprise as we know the
$(2,1)$-component to preserve supersymmetry.

\vskip0.2cm
\noindent
{\it (ii) $(0,3)$-flux}
\vskip0.2cm

The $(0,3)$-flux is imaginary self-dual as well. It is the only flux component with 
$\hat{W}\neq 0$. This leads to non-zero $F^{T^i}$ and non-zero gravitino mass
$m_{3/2}=e^{\kappa_4^2\ \hat{K}/2}\kappa_4^2\ \hat{W}$. The scalar
potential in the bulk vanishes, therefore
\eqn\msofti{\eqalign{
(m^{33}_{i\ov i})^2&=\kappa_4^2\ \lf(|m_{3/2}|^2G_{C^3_i\ov
C^3_i}-|F^{T^i}|^2R^{3}_{T^i\ov T^ii\ov{i}}\ri),\cr
(m^{7,j}_{i\ov i})^2&=\kappa_4^2\ \lf(|m_{3/2}|^2G_{C^{7,j}_i\ov
C^{7,j}_i}-\sum_{k,l}F^{T^k}\ov F^{\ov T^l}R^{7,j}_{T^k\ov T^l i\ov i}\ri),\cr
(m^{37_a})^2&=\kappa_4^2\ \lf(|m_{3/2}|^2G_{C^{37_a}\ov
C^{37_a}}-\sum_{k,l}F^{T^k}\ov F^{\ov T^l}R^{37_a}_{T^k\ov T^l}\ri),\cr
(m^{7a7b})^2&=\kappa_4^2\ \lf(|m_{3/2}|^2G_{C^{7a7b}\ov
C^{7a7b}}-\sum_{k,l}F^{T^k}\ov F^{\ov T^l}R^{7a7b}_{T^k\ov T^l}\ri)\ ,}}
The scalar mass term for the 3-brane deserves our special attention. After
inspection of our findings of appendix \appB, we find that $|F^{T^i}|^2R^{3}_{T^i\ov
T^ii\ov{i}}=|m_{3/2}|^2G_{C^3_i\ov C^3_i}$, therefore
$$(m^{33}_{i\ov i})^2=0.$$
The same would be true for the scalars on the 7-branes without 2-form
fluxes. Turning on 2-form flux destroys the simple structure of the curvature
tensors and thus generates scalar mass terms.

The trilinear coupling becomes
\eqn\aiik{\eqalign{
A_{IJK}&=3i\ Y_{IJK}\prod_M(M-\ov M)^{-1}{\lambda\kappa_4^2\over
(2\pi)^2\alpha'}\int \ov G_3\wedge \ov\Om\cr
&-i\prod_M(M-\ov M)^{-1/2}\sum_{i=1}^3F^{T^i} G^{\ov C_IC_I}\partial_{T^i}
G_{C_I(\ov C_I}Y_{JK)I}.}}
For $I,\ J,\ K$ referring to $C^3_i$, the trilinear coupling is zero.
The gaugino masses are
$$m_{g,
D7,j}={F^{T^j}\over{(T^j-\ov{T}^j)-\alpha'^2f^kf^l(S-\ov{S})}}\ .$$
To have an explicit example, we will substitute our solution for the
$(0,3)$-flux \eqnulldrei\  for $U^1=U^2=U^3=S=i$ into the above formulas. The superpotential is
$\hat{W}=8\lambda(d_3-id_0)$.
The $F$-terms become
$$F^{T^i}=-2\lambda\ \kappa_4^2\ (d_3+id_0)(T^i-\ov{T}^i)^{1/2}(T^j-\ov{T}^j)^{-1/2}
(T^k-\ov{T}^k)^{-1/2},\quad i\neq j\neq k.$$ 
The gravitino mass is
$$|m_{3/2}|^2=4\lambda^2\ \kappa_4^4\ \prod_i|T^i-\ov T^i|^{-1}(d_3^2+d_0^2).$$
The gaugino masses are
$$m_{g, D7,j}=-2\lambda\ \kappa_4^2\ (d_3+id_0){(T^j-\ov{T}^j)^{1/2}(T^k-\ov{T}^k)^{-1/2}
(T^l-\ov{T}^l)^{-1/2}\over{(T^j-\ov{T}^j)-2i\alpha'^2f^kf^l}}\ .$$

\vskip0.2cm
\noindent
{\it (iii) $(1,2)$-flux}
\vskip0.2cm
Here, only $F^{U^i}\neq 0$. The gravitino mass $m_{3/2}$ vanishes again and
$m_g=0$. For the remaining terms, we find
\eqn\softeinszwei{\eqalign
{\hat{V}&=-\kappa_4^{-2}\ \sum_{i=1}^3{{1}\over{(U^i-\ov{U}^i)^2}}|F^{U^i}|^2,\cr
(m^{33}_{i\ov i})^2&=\kappa_4^2\ \left[\kappa_4^2\hat{V}G_{C^3_i\ov
C^3_i}-|F^{U^i}|^2R^{3}_{U^i\ov U^ii\ov{i}}\right],\cr
(m^{7,j}_{i\ov
i})^2&=\kappa_4^2\ [\kappa_4^2\hat{V}G_{C^{7,j}_i\ov
C^{7,j}_i}-\sum_{k,l}F^{U^k}\ov F^{\ov U^l}R^{7,j}_{U^k\ov U^l i\ov i}],\cr
(m^{37_a})^2&=\kappa_4^2\ [\kappa_4^2\hat{V}G_{C^{37_a}\ov
C^{37_a}}-\sum_{k,l}F^{U^k}\ov F^{\ov U^l}R^{37_a}_{U^k\ov U^l}],\cr
(m^{7a7b})^2&=\kappa_4^2\ [\kappa_4^2\hat{V}G_{C^{7a7b}\ov
C^{7a7b}}-\sum_{k,l}F^{U^k}\ov F^{\ov U^l}R^{7a7b}_{U^k\ov U^l}],\cr
A_{IJK}&=i\prod_M(M-\ov M)^{-1}{\kappa_4^2\lambda\over
(2\pi)^2\alpha'}\sum_i\int \ov G_3\wedge \ov\om_{A_i}(Y_{IJK}-(U^i-\ov U^i)\p_{U^i}Y_{IJK})\cr
&-i\prod_M(M-\ov M)^{-1/2}\sum_{i=1}^3F^{U^i} G^{\ov
C_IC_I}\partial_{U^i} G_{C_I(\ov C_I}Y_{JK)I},}}
which is again zero for the case of $I,\ J,\ K$ referring to $C^3_i$.
We will now substitute our solution \eqeinszwei \ as an example. Here, we have
\eqn\omeinszwei{\eqalign{
{1\over{(2\pi)^2\alpha'}}\ \int\ov G_3\wedge
\omega_{A1}&=-4\ (d_2+d_3+ib_2+ib_3),\cr
{1\over{(2\pi)^2\alpha'}}\ \int\ov G_3\wedge
\omega_{A2}&=4\ (-d_1-d_3+ib_2+id_0),\cr
{1\over{(2\pi)^2\alpha'}}\ \int\ov G_3\wedge
\omega_{A3}&=4\ (-d_1-d_2+ib_3+id_0).}}
For the $F$-terms, we get the following expressions:
\eqn\Feinszwei{\eqalign{
F^{U^1}&=2i\lambda\kappa_4^2\prod(T^i-\ov T^i)^{-1/2}(-d_2-d_3+ib_2+ib_3),\cr
F^{U^2}&=-2i\lambda\kappa_4^2\prod(T^i-\ov T^i)^{-1/2}(d_1+d_2+ib_2+id_0),\cr
F^{U^3}&=-2i\lambda\kappa_4^2\prod(T^i-\ov T^i)^{-1/2}(d_1+d_2+ib_3+id_0).}}
So together with what we learned from eq. \VhutTc , $\hat V$ is
$$\hat V=4\lambda^2\kappa_4^2\prod_i|T^i-\ov T^i|^{-1}\sum|B^i|^2.$$

\vskip0.2cm
\noindent
{\it (iv) $(3,0)$-flux}
\vskip0.2cm

In this case, the only non-vanishing $F$-term is $F^S$, this corresponds to
the ``dilaton domination'' SUSY breaking. The scalar potential in the bulk is
$$\hat{V}=-{\kappa_4^{-2}\over{(S-\ov{S})^2}}|F^S|^2.$$
The gravitino mass $m_{3/2}$ is zero,
\eqn\softdreinull{\eqalign{
(m^{33}_{i\ov i})^2&=\kappa_4^4\hat{V}G_{C^3_i\ov
C^3_i},\cr
(m^{7,j}_{i\ov
i})^2&=\kappa_4^2\ [\kappa_4^2\hat{V}G_{C^{7,j}_i\ov
C^{7,j}_i}-|F^S|^2R^{7,j}_{S\ov S i\ov i}],\cr
(m^{37_a})^2&=\kappa_4^2\ [\kappa_4^2\hat{V}G_{C^{37_a}\ov
C^{37_a}}-|F^S|^2R^{37_a}_{S\ov S}],\cr
(m^{7a7b})^2&=\kappa_4^2\ [\kappa_4^2\hat{V}G_{C^{7a7b}\ov
C^{7a7b}}-|F^S|^2R^{7a7b}_{S\ov S}],\cr
A_{IJK}&=i\ Y_{IJK}\prod_M(M-\ov M)^{-1}{\kappa_4^2\lambda\over
(2\pi)^2\alpha'}\int G_3\wedge \ov\Om\cr
&-i\prod_M(M-\ov M)^{-1/2}F^S G^{\ov C_IC_I}\partial_S G_{C_I(\ov
C_I}Y_{JK)I},\cr
m_{g,
D7,j}&=F^S{{-\alpha'^2f^kf^l}\over{(T^j-\ov{T}^j)-\alpha'^2f^kf^l(S-\ov{S})}},\cr
m_{g, D3}&=F^S{1\over(S-\ov S)}=-i\prod_M(M-\ov
M)^{-1/2}{\lambda\kappa_4^2\over (2\pi)^2\alpha'}\int G_3\wedge \ov\Om.
}}
We will now substitute our solution \eqdreinull \ as an example. Here, we have
${1\over{(2\pi)^2\alpha'}}\int\ov G_3\wedge \Om=8(d_3-id_0)$. With this, $F^S$ becomes
$$F^S=4i\lambda\kappa_4^2\ (d_3+id_0)\prod_i(T^i-\ov T^i)^{-1/2}. $$
So $\hat V$ is
$$\hat V=4\lambda^2\kappa_4^2\ (d_3^2+d_0^2)\prod_i|T^i-\ov T^i|^{-1},$$
and finally, 
\eqn\gadn{\eqalign{
m_{g, D7j}&=4i\lambda\kappa_4^2\ (d_3+id_0)\prod_i(T^i-\ov
T^i)^{-1/2}{{-\alpha'^2f^kf^l}\over{(T^j-\ov{T}^j)-2i\alpha'^2f^kf^l}},\cr
m_{g, D3}&=2\lambda\kappa_4^2\ (d_3+id_0)\prod_i(T^i-\ov
T^i)^{-1/2}.}}


\newsec{$F$--theory description}

The description so far is valid as long as we are in the regime of small string coupling 
constant $g_s\sim e^{1/2\phi_{10}}$ and string tension $\ap\sim M_{string}^{-2}$ 
(or large radius limit).
As we have seen, flux--quantization usually fixes the dilaton at finite values
$g_s\sim 1$.
Let us emphasize, that in the orientifold construction introduced in section 2, 
the localized tadpoles of both the $O3$-- and $O7$--planes are cancelled in a non--local way.
This quite generically produces a non--constant dilaton. This means, that the supergravity 
equations of motion lead to a dependence of the dilaton $\tau$ over the internal compact manifold.
The natural setup to describe compactifications with a dilaton varying over the 
compactification manifold is $F$--theory compacified on elliptically fibered fourfolds.
In this section, we shall lift our \tb orientfold construction to $F$--theory.
This allows a more thorough discussion of the supersymmetry breaking effects discussed before.
Moreover, it provides the non--perturbative formulation of the \tb orientfold constructions
with background fluxes.

We shall discuss $F$--theory compactified on a fourfold $X_8$ \doubref\CV\BJPS. The latter
is an elliptic fibration over a threefold base $X_6$, to be specified later.
This compactification gives rise to N=1 supersymmetry in $D=4$ and the low--energy properties
are determined by the fourfold, the $D3$-- and $D7$--brane configurations. 
The complex structure modulus of the fiber, being a non--trivial function on the base
coordinates, represents the complex coupling constant.
The fluxes $G_3\in H^{3}(X_6,\IZ)$ of the \tb superstring become elements of the integer 
cohomology $G_4\in H^{4}(X_8,\IZ)$ of the manifold $X_8$.
In $F$--theory compactified on the fourfold $X_8$ with background fluxes $G_4$, 
the tadpole condition \modcfour\ becomes \DRS:
\eqn\Fcfour{
N_{D3}=\fc{1}{24}\ \chi(X_8)-\fc{1}{8 \pi^2}\int_{X_8} G_4\wedge G_4-\sum_{a=1}^{N_{D7}}
\int_{C_{a;4}} F\wedge F\ .}
We have $N_{D7}$ $D7$--branes wrapped around $4$--cycles of the base manifold $X_6$.
The last term accounts for the total number of instantons inside the $D7$--branes, which
arise from background gauge bundles on the compact part of the $D7$--brane world--volume. 
The latter break the gauge group down from the maximal one, dictated by
the singularity of the elliptic fibration. W.r.t. the base manifold $X_6$, the singularity 
is a complex codimension 2 locus, around which the $D7$--brane is wrapped.
The term proportional to the Euler number $\chi(X_8)$ of
the fourfold $X_8$ accounts for the total induced $D3$--brane charge 
coming from wrapping the $D7$--branes over $4$--cycles of the base $X_6$ \SVW.
The Euler number $\chi(X_8)$ of the fourfold $X_8$ is given by:
\eqn\euler{
\chi(X_8)=\int\limits_{X_8}I_8(R)\ \ \ ,\ \ \ I_8(R)=\fc{1}{192}\ \lf[\ \tr R^4-\fc{1}{4}\ 
(\tr R^2)^2\ \ri]\ .}
The $F$--theory lift of the superpotential \TVW\ becomes \GVW:
\eqn\Fsup{
\hat W=\int_{X_8} \Omega_4\wedge G_4\ .}

In the following, we shall consider the fourfold
\eqn\fourfold{
X_8=\fc{T^6/\Gamma \times T^2}{\IZ_2}\ ,}
with $\Gamma=\IZ_2\times \IZ_2$ being the group \generator\ introduced in section 2.
A third $\IZ_2$ generator accounts for the elliptic fibration over our three--fold 
base $X_6=T^6/\Gamma$, discussed in the previous sections.
More concretely, we have the three $\IZ_2$--actions
\eqn\generatorF{\eqalign{
\th_1\ :\ & (z^1,z^2,z^3,z^4)\lra (-z^1,-z^2,z^3,z^4)\ ,\cr
\th_2\ :\ & (z^1,z^2,z^3,z^4)\lra (z^1,-z^2,-z^3,z^4)\ ,\cr
\th_3\ :\ & (z^1,z^2,z^3,z^4)\lra (z^1,z^2,-z^3,-z^4)}}
acting on the four internal complex coordinates $z^i\ ,\ i=1,2,3,4$.
The two generators $\th_1$ and $\th_2$ are the generators \generator.
The third generator $\th_3$ reflects both the coordinates $z^3$ and $z^4$. 
The latter is the complex coordinate of the elliptic fiber $T^2$.

Let us briefly discuss the orientifold limit of this $F$--theory compactification, \ie
how our previously studied $\Om(-1)^{F_L}I_6$--orientifold of $T^6/\IZ_2\times \IZ_2$ arises.
Reflecting the fiber coordinate $z^4\ra -z^4$ corresponds to the monodromy element $-{\bf 1}_2\in 
SL(2,\IZ)$, changing the sign of the two $2$--form fields $B_2$ and
$C_2$. However, it leaves  all other massless fields invariant. It represents a perturbative 
symmetry equivalent to the orientifold action $\Om(-1)^{F_L}$ \SEN.
Hence, the generator $\th_3$ corresponds to the orientifold group element 
$\th_3 \equiv \Om(-1)^{F_L}I_2^3$, discussed in section 2.
Furthermore, we have the following identifications:
$\th_2\th_3\equiv \Om(-1)^{F_L}I_2^2\ ,\ 
\th_1\th_2\th_3\equiv \Om(-1)^{F_L}I_2^1$ 
and $\th_1\th_3\equiv \Om(-1)^{F_L}I_6$.
The positions of the $D7$--branes and $O7$--planes on the base $X_6$ are given by the zeros 
of the discriminant of the elliptic fiber in an appropriate constant coupling limit.

The manifold $(T^2)^4/\IZ_2^3$ belongs to the family of Borcea fourfolds.
The latter are singular Calabi--Yau fourfolds with $SU(4)$ holonomy, 
defined by the quotient $(K_3\times K_3)/\IZ_2$.
Here, the $\IZ_2$ acts as an involution, described by $(r_1,a_1,\delta_1)$ on the first $K3$,  
and as an involution, described by $(r_2,a_2,\delta_2)$ on the second $K3$. It
reverses the sign of the $(2,0)$--forms of each $K_3$, but leaves the $(4,0)$--form of
the fourfold $X_8$ invariant.
Their Euler numbers are given by $\chi=288+6\ (r_1-10)(r_2-10)$.
The fourfold \fourfold\ under consideration corresponds\foot{Other choices
would correspond to \tb orientifolds with discrete torsion.} to the case\foot{Note, that
the orbifold $X_6=T^6/(\IZ_2\times\IZ_2)$ itself corresponds to an elliptic fibration
of Voison-Borcea type $(K_3\times T^2)/\IZ_2$, with the $\IZ_2$--action 
decribed by the element $\th_2$. The latter reverses the sign of the $(1,0)$--form of the 
$T^2$ and the sign of the $(2,0)$--form of the $K_3$. 
The numbers $(r_2,a_2,\delta_2)=(18,4,0)$ encode the $CY$ data
$h_{(1,1)}(X_6)=51$ and $h_{(2,1)}(X_6)=3$, provided the $K3$ is realized as $T^4/\IZ_2$ orbifold
limit.}, 
with $r_1=r_2=18$,\ $a_1=a_2=4$, and $\delta_1=\delta_2=0$.
Hence for the manifold \fourfold\ we have:
\eqn\EULER{
\chi(X_8)=672\ .}
With this information, the tadpole condition \Fcfour\ becomes: 
\eqn\Fcfoursp{
N_{D3}=28-\fc{1}{8 \pi^2}\int_{X_8} G_4\wedge G_4-\sum_{a=1}^{N_{D7}}
\int_{C_{a;4}} F\wedge F\ .}
Without $4$--form fluxes, \ie $G_4=0$, this equation reduces to:
\eqn\withoutflux{
N_{D3}=28-\sum_{a=1}^{N_{D7}}\int_{C_{a;4}} F\wedge F\ .}
This equation essentially describes the tadpole condition \cfour\ of the \tb orientifold
discussed in the previous sections. Without instanton bundles on the $D7$--branes
we would conclude, that $28$ $D3$--branes are necessary to cancel the Ramond $4$--form charge.
However, some care is necessary for the interpretation 
of \eqq \withoutflux\ \kakui:
$24$ of the $28$ $D3$--branes are dissolved into instantons on the $96$ $D7$--branes, leaving
us with $4$ $D3$--branes. The latter are placed on the orientifold $O3$--planes and
each $D3$--brane appears in an orbit of $8$ due to the group elements 
$\th_1,\th_2$ and $\Om$. 
Hence effectively, we have $4\times 8=32$ $D3$--branes in agreement with the model presented
after \eqq \susyi.

A general $4$--flux $G_4$ on $X_8$ may be expressed as the sum 
$G_4=G_{(4,0)}+G_{(3,1)}+G_{(2,2)}+G_{(1,3)}+G_{(0,4)}$, 
written as a linear combination of elements
of $H^{(p,q)}(X_8,\IC)$, with $p+q=4$. 
Since $h_{(4,0)}(X_8)=h_{(0,4)}(X_8)=1$, we have the components:
\eqn\allowed{
dz^1\wedge dz^2\wedge dz^3\wedge dz^4\ \ \ ,\ \ \ d\ov z^1\wedge d\ov z^2\wedge d\ov z^3\wedge 
d\ov z^4\ .}
Besides, due to $h_{(3,1)}(X_8)=4$, we have the $(3,1)$--components:
\eqn\allowedi{\eqalign{
& d\ov z^1\wedge dz^2\wedge d z^3\wedge d z^4\ \ \ ,\ \ \ 
  dz^1\wedge d\ov z^2\wedge dz^3\wedge d z^4\cr
& dz^1\wedge dz^2\wedge d\ov z^3\wedge d z^4\ \ \ ,\ \ \ 
  dz^1\wedge dz^2\wedge dz^3\wedge d\ov z^4\ .}}
Similarily, for the $(1,3)$--components, with $h_{(1,3)}(X_8)=4$.
Finally, due to $h_{(2,2)}(X_8)=460$, we have many $(2,2)$ components 
from the untwisted and twisted sector.
The untwisted sector gives rise to the six $(2,2)$--components:
\eqn\allowedii{\eqalign{
&d z^1\wedge dz^2\wedge d \ov z^3\wedge d \ov z^4\ \ \ ,\ \ \ 
d z^1\wedge d\ov z^2\wedge d z^3\wedge d \ov z^4
\ \ ,\ \ d \ov z^1\wedge dz^2\wedge d z^3\wedge d \ov z^4\cr
&d z^1\wedge d\ov z^2\wedge d \ov z^3\wedge d  z^4\ \ \ ,\ \ \ 
d \ov z^1\wedge dz^2\wedge d \ov z^3\wedge d z^4
\ \ ,\ \ d \ov z^1\wedge d \ov z^2\wedge d z^3\wedge d z^4\ .}}
In the following, we shall not discuss the remaining $454$ $(2,2)$--forms
corresponding to flux components from the twisted sector. It is 
assumed that their quantization rules freeze the blowing up moduli of the orbifold \fourfold.
All the fluxes $G_{(p,q)}$, displayed above, are invariant under the orbifold group \generatorF.
In addition, they are primitive\foot{At any rate, on a fourfold, all components
are primitive, except $G_{(2,2)}$. The latter may have primitive components being self--dual
and non--primitive ones being anti--selfdual.}, \ie they fulfill $J^{5-p-q}\wedge G_{(p,q)}=0$,
with the K\"ahler form $J$ on $X_8$: $J=i\sum\limits_{i=1}^4 \Tc_2^i\ dz^i\wedge d\ov z^i$.
Here, $\Tc_2^4$ is the K\"ahler modulus of the fiber torus. 
Due to $\star F_{(p,4-p)}=(-1)^p\ F_{(p,4-p)}$, which holds for primitive flux components \GVW, 
the fluxes \allowedi\ are anti--selfdual. The remaining fluxes \allowed\ and \allowedii\
are self--dual.
In the case of unbroken supersymmetry, the $4$--form fluxes $G_4$ have to be 
self--dual elements $G_{(2,2)}$ \BB, leading to a vanishing superpotential \Fsup.
On the other hand, the components $G_{(4,0)}$ and $G_{(0,4)}$ break supersymmetry with a 
vanishing scalar potential.
Only anti--selfdual components of $G_4$, \ie $G_{(3,1)}$ and $G_{(1,3)}$, 
lead to a non--vanishing scalar potential \doubref\GVW\HL. 
To this end, in our case, a non--vanishing scalar potential arises only from
non--vanishing components $G_{(3,1)}$ and $G_{(1,3)}$.

The supersymmetry preserving $(2,1)$--form fluxes $G_3$, 
found in subsection 2.2 for our \tb orientifold 
compactification on $X_6$, may be directly lifted to self--dual 
$(2,2)$--form fluxes $G_4$ of $F$--theory. 
Lorentz--invariance in $D=4$ demands, that one component of the four--flux $G_4$
refers to the elliptic fiber. The other three components refer to the base manifold $X_6$ \DRS.
Another argument is, that a self--dual integer $4$--form flux may describe
an $F$--theory limit only if it has one of its components w.r.t. the fiber \GVW.
Hence for that case, the $4$--form flux may be written in the $SL(2,\IZ)_S$--invariant
combination:
\eqn\ansatzfourflux{
G_4=\fc{2\pi}{(S-\ov S)}\ (\ G_3\wedge d\ov z^4-\ov G_3\wedge dz^4\ )\ .}
This form of the $4$--flux is at least appropriate in a local trivialization of the elliptic
fibration \GKP.

In the following, let us make contact to our findings from subsection 2.3.
When we insert $dz^4=dx^4+Sdy^4$ into the above equation, we find
\eqn\gfour{
G_4=-2\pi\ [\ F_3\wedge dy^4-H_3\wedge dx^4\ ]\ .}
We now want to express $G_4$ through the language of section 2.3. To do so, we
first define our basis for the 4-form flux:
\eqn\Frealbase{\eqalign{
\tilde \alpha_0&= \alpha_0\wedge dx^4,\quad
\tilde \alpha_i= \alpha_i\wedge dy^4,\ i=1,\ldots,3,\cr
\tilde \gamma^0&= \beta_0\wedge dy^4,\quad
\tilde \gamma^i= \beta_i\wedge dx^4,\ i=1,\ldots,3,\cr
\tilde \beta_0&= \alpha_0\wedge dy^4,\quad
\tilde \beta_i= \alpha_i\wedge dx^4,\ i=1,\ldots,3\cr
\tilde \delta^0&= \beta_0\wedge dx^4,\quad
\tilde \delta^i= \beta_i\wedge dy^4\ i=1,\ldots,3,}}
where the $\alpha,\ \gamma$ correspond to ISD fluxes and the $\beta,\ \delta$
to IASD fluxes. Expressed in this basis, our lifted 3-form flux becomes
\eqn\realfour{\eqalign{
{-1\over(2\pi)^3\alpha'}G_4&=c^0\tilde\alpha_0+\sum_{i=1}^3a^i\tilde\alpha_i+
\sum_{i=1}^3d_i\tilde\gamma^i+b_0i\tilde\gamma^0\cr
&+a^0\tilde\beta_0+\sum_{i=1}^3c^i\tilde\beta_i+\sum_{i=1}^3b_i\tilde\delta^i+d_0\tilde\delta^0,}}
with the $a,\ b,\ c,\ d\ $ the coefficients of our original 3-form flux.
Now we want to do the same in complex notation. We define the following
complex basis:
\eqn\Fcplxbase{\eqalign{
\tilde\om_{A0}&=\om_{B0}\wedge dz^4,\quad \tilde\om_{Ai}=\om_{Bi}\wedge
dz^4,\cr
\tilde\om_{B0}&=\om_{B0}\wedge d\bar z^4,\quad \tilde\om_{Bi}=\om_{Ai}\wedge
dz^4,\cr 
\tilde\om_{C0}&=\om_{A0}\wedge d\bar z^4,\quad \tilde\om_{Ci}=\om_{Ai}\wedge d\bar
z^4,\cr
\tilde\om_{D0}&=\om_{A0}\wedge dz^4,\quad \tilde\om_{Di}=\om_{Bi}\wedge d\bar
z^4,\cr
}}
where the $\om_{A},\ \om_{C}$ correspond to the ISD fluxes and the  $\om_{B},\
\om_{D}$ correspond to the IASD fluxes, and $\ov{\tilde\om_A}=\tilde\om_C,\
\ov{\tilde\om_B}=\tilde\om_D$.
Expressed through the $\,\tilde\alpha,\ \tilde\beta,\ \tilde\gamma,\
\tilde\delta$ and the $U^i$ and 
$S$, $\ \tilde \om_{A_0}$ for example has the following form:
\eqn\omtilde{\tilde\om_{A0}=\tilde\alpha_0+S\tilde\beta_0+\sum_{i=1}^3\ov
U^i(\tilde\beta_i+S\tilde\alpha_i)-\sum_{i\neq j\neq k}\ov U^j\ov
U^k(\tilde\gamma^i+S\tilde\delta^i)+\ov U^1\ov U^2\ov U^3(\tilde\delta^0+S\tilde\gamma^0).}
The form of the other $\om$ can be easily deduced from the above and
\cplxbase.
This basis fulfills
\eqn\baseprop{\eqalign{
\int\om_{Ai}\wedge\om_{Ci}=-\int \om_{Bi}\wedge\om_{Di}=-\prod_{j=1}^3(U^j-\ov
U^j)(S-\ov S),\ i=0,\ldots,3}}
while all other combinations of the basis elements are zero.
Expressed through this complex basis, our 4-form flux \ansatzfourflux\  has the
following form:
\eqn\cplxfour{\eqalign{
{(S-\ov S)\over
(2\pi)^3\alpha'}\ G_4&=A^0\tilde\om_{C0}+\sum_{i=1}^3A^i\tilde\om_{Ci}
+B^0\tilde\om_{B0}+\sum_{i=1}^3B^i\tilde\om_{Di}\cr
&-\ov A^0\tilde\om_{A0}-\sum_{i=1}^3\ov A^i\tilde\om_{Ai}-\ov
B^0\tilde\om_{D_0}-\sum_{i=1}^3\ov B^i\tilde\om_{Bi}\ ,}} 
where again the $A^i,\ B^i$ are the complex coefficients of our original
3-form flux. 
With this, we find: 
\eqn\explFflux{
N_{flux}=-{1\over 8\pi^2}\int G_4\wedge G_4=-\ {\prod\limits_{i=1}^3(U^i-\ov
U^i)\over(S-\ov S)}\ \lf(\sum_{i=0}^3|A^i|^2-\sum_{i=0}^3|B^i|^2\ri)\ ,}
which agrees with $N_{flux}$ from \eqq \cplxNflux.
The anomaly equation \Fcfoursp\ 
becomes:
\eqn\anomaly{
N_{D3}=28+N_{flux}-\sum_{a=1}^{N_{D7}}\int_{C_{a;4}} F\wedge F\ .}

So far, we have described the $F$--theory lift of our \tb orientifold construction
with $3$--form fluxes. 
Generically, $F$--theory compactified on Calabi--Yau fourfolds with $G_4$--fluxes leads 
to a warped metric.
See {\it Refs.} \fiveref\GVW\DRS\verlinde\PM\greene\ for an account on this subject.
The next step would be to work out the differential equation for
the warp factor following from the equations of motion for the $4$--form field $C_4$.
Furthermore, from the equation of motion for the internal metric and dilaton field, 
a system of differential equations follows for the (non--constant) dilaton field.
We reserve this for future work.

\newsec{Concluding remarks}

In this article, we have determined the $D=4, N=1$ tree--level action for a class of 
\tb orientifold models with $D3$-- and $D7$--branes up to second order in the matter fields.
These models are based on toroidal orbifold/orientifold compactifications of type $IIB$.
The action, summarized in the three functions $K,W$ and $f$, 
depends on both closed and open string moduli fields.
Generically, the closed string moduli fields describe the geometry of the underlying 
\tb compactification and the open string moduli account for moduli fields originating from the
$D$--branes like matter fields, Wilson lines and the $D$--brane positions.
We have calculated the K\"ahler metrics of these moduli fields extending the results 
of {\it Ref.}~\LMRS. In particular, in subsection 3.3. the metric for matter fields originating
from a $1/2$ BPS system of two $D$--branes is elaborated upon.
We have allowed for both non--vanishing $RR$ and $NSNS$ $3$--form fluxes in the bulk and 
$2$--form fluxes (instantons) on the $D7$--brane world--volume. 
While the former allow for fixing the dilaton and 
complex structure moduli due to their internal flux quantization conditions, 
non--vanishing $2$--form fluxes give rise to a $D$--term potential fixing a part of 
the K\"ahler moduli.
The K\"ahler potential $K$ and the gauge kinetic function $f$ only depend non--trivially on
the $2$--form fluxes (\cf \eqqs \KW\ and \gauge). On the other hand, as it is well--known, 
the $3$--form fluxes enter the holomorphic superpotential $W$ (\cf \TVW).
The latter depends on the dilaton and complex structure moduli
only. A fact, which is at least true in the no--scale case. However, since the calculation of
the scalar potential and possible soft--supersymmetry breaking terms involves 
both the K\"ahler potential, gauge kinetic function
and the superpotential, those terms depend non--trivially both on $2$-- and $3$--form 
fluxes (\cf \eqqs \msoft\ and \Aijk).

Generically, most of the discussions on superstring vacua with background fluxes
is at the level of the lowest order expansion in the string coupling $g_s$ and string tension
$\ap$ of the underlying superstring effective action.
Hence the stabilization of some moduli takes place at string tree--level. In contrast, one
should mention, that there exist purely {\it stringy} constructions like $M$--theory $U$--duality 
orbifolds with a large number of moduli frozen by the orbifold group and 
with very few moduli left unfixed \Muduality.
In type $IIB$, the superpotential is exact to all orders in $\ap$ as it only depends on
the dilaton and complex structure moduli and does not receive world--sheet instanton corrections.
However, through $\ap$--corrections
to the K\"ahler potential, eventually also the $F$--terms and the scalar potential 
receive corrections. Hence, in general, these effects have to be taken into account
and it is certainly desirable to go beyond the lowest order approximation.
Already if one includes one--loop effects into the K\"ahler potential, the no--scale structure is 
generically lost \LH. In addition, the no--go theorem of supergravity, that only
a certain class of fluxes (namely ISD--fluxes) lead to a consistent supergravity solution
may be disproved by some additional stringy effects.
Hence the study of effects, which go beyond the supergravity approximation, is very important.

\bigskip
\centerline{\bf Acknowledgments }\nobreak
\bigskip
We would like to thank C. Angelantonj, R. Blumenhagen and P. Mayr for valuable discussions.
This work is supported in part by the Deutsche 
Forschungsgemeinschaft (DFG), and the German--Israeli Foundation (GIF).

\appendix\appA{Fluxes in the presence of $D3$ and $D7$--branes}

One way one can think of turning on flux on a $Dp$-brane is via the generalized 
Scherk-Schwarz Ansatz:
\eqn\SchSch{B_{mn}=H_{mnp}x^p.}
$p$ may run only over the coordinates transversal to the brane, so in the case
of a $D7$-brane, which fills the directions $x_0,\ldots,x_7$ (wrapping the
tori $T_1^2\times T_2^2$), we have
\eqn\SSseven{B_{mn}^7=H_{mn8}x^8+H_{mn9}x^9.}
As $H_{ijk}$ must always have one index equal to either 8 or 9, not all of the
20 possible components are allowed in our case. Not allowed are the fluxes

\eqn\notallowed{
\eqalign{&dx^1 \wedge dy^1  \wedge dx^2 \qquad dx^1 \wedge dy^1 \wedge dy^2 \cr
&dx^1 \wedge dx^2  \wedge dy^2 \qquad dy^1 \wedge dx^2
 \wedge dy^2.}}
Expressed in complex notation, this would correspond to the fluxes
$H_{1\ov{1}2},\ H_{1\ov{2}2},\ H_{1\ov{1}\ov{2}},\ H_{2\ov{1}\ov{2}}$.

For $D7$-branes wrapping the tori $T_1^2\times T_3^2$ or $T_2^2\times T_3^2$,
we find similar results. Having a setup of three stacks of $D7$ branes, one
stack not wrapping $T_3^2$, one not wrapping  $T_2^2$, and one not wrapping
$T_1^2$, we lose 12 of the twenty flux components and are left with fluxes,
which have one index on each of the tori.

\appendix\appB{Components of the curvature tensor}

We will first derive the components coming from the metric for the matter
fields on the 3-branes.
Due to the diagonal structure of the metric, many components of the curvature
tensor are zero. We are thus left only with the following non-zero components:
\eqn\curvthree{\eqalign{
R^3_{T^i\ov T^ii\ov i}&={-1\over{(T^i-\ov T^i)^2}}G_{C^3_i\ov C^3_i},\cr
R^3_{U^i\ov U^ii\ov i}&={-1\over{(U^i-\ov U^i)^2}}G_{C^3_i\ov C^3_i}.}}
Now we will examine the components coming from the metric for matter fields on the stack of 7-branes not wrapping the
$j$'th torus. If we assume the 2-form flux to be zero, we arrive at the same 
component structure as for the 3-brane. Turning on the 2-form flux makes life a lot
harder, as the $\tilde f^i={f^i \over {\rm Im}{\cal T}^i}$ depend on all of the
$T^i$ and on $S$. Only few components are zero now. The only one that retains
the simple structure from above is
\eqn\RUU{R^{7,j}_{U^iU^ii\ov{i}}={-1\over{(U^i-\ov U^i)^2}}G_{C^{7,j}_i\ov C^{7,j}_i}.}
For the other components, we need to know the $\partial_M({\rm Im}{\cal
T}^j)$, where $M$ runs over $S,\ T^i$. They can be obtained by taking the
total derivative of the imaginary part of equations \fieldT \ and \fieldS \
and inverting the partial derivatives. We find
\eqn\deriv{\eqalign{
{\partial({\rm Im}{\cal T}^j)\over \partial S}&={-i\over4}{{\rm Im}{\cal T}^j\over
{\rm Im}S},\cr
{\partial({\rm Im}{\cal T}^j)\over \partial T^j}&={-i\over4}{{\rm Im}{\cal T}^j\over
{\rm Im}T^j},\cr
{\partial({\rm Im}{\cal T}^j)\over \partial T^k}&={i\over4}{{\rm Im}{\cal T}^j\over
{\rm Im}T^k},\ j\neq k.\cr
}}
We will also need the $\partial_M\partial_{\ov N}({\rm Im}{\cal T}^j)$:
\eqn\doublederiv{\eqalign{
{\partial({\rm Im}{\cal T}^j)\over {\partial S \partial \ov S}}&=
{-3\over 16}{({\rm Im}{\cal T}^j)\over({\rm Im}S)^2}, \qquad
{\partial({\rm Im}{\cal T}^j)\over {\partial T^j \partial \ov S}}=
{-1\over 16}{({\rm Im}{\cal T}^j)\over({\rm Im}S)({\rm Im}T^j)},\cr
{\partial({\rm Im}{\cal T}^j)\over {\partial T^k \partial \ov S}}&=
{1\over 16}{({\rm Im}{\cal T}^j)\over({\rm Im}S)({\rm Im}T^k)},\quad
{\partial({\rm Im}{\cal T}^j)\over {\partial T^k \partial \ov T^i}}=
{-1\over 16}{({\rm Im}{\cal T}^j)\over({\rm Im}T^i)({\rm Im}T^k)},\cr
{\partial({\rm Im}{\cal T}^j)\over {\partial T^k \partial \ov T^j}}&=
{1\over 16}{({\rm Im}{\cal T}^j)\over({\rm Im}T^j)({\rm Im}T^k)},\quad
{\partial({\rm Im}{\cal T}^j)\over {\partial T^k \partial \ov T^k}}=
{1\over 16}{({\rm Im}{\cal T}^j)\over({\rm Im}T^k)^2},\cr
{\partial({\rm Im}{\cal T}^j)\over {\partial T^j \partial \ov T^j}}&=
{-3\over 16}{({\rm Im}{\cal T}^j)\over({\rm Im}T^j)^2},
}}
where $M,\ N$ run over $S,\ T^i$.
Now we can write down the remaining $R^{7,j}$:
\eqn\Rmixed{\eqalign{
R^{7,j}_{M \ov N i\ov{i}}&={-\kappa_4^{-2}\over{(T^k-\ov{T}^k)(U^i-\ov
U^i)}}{1\over{|1+i\tilde{f}^i|^3|1+i\tilde{f}^k|}}\times\cr
&\times\left[|1+i\tilde{f}^k|^2\p_M|1+i\tilde{f}^i|\p_{\ov N}|1+i\tilde{f}^i|-\right.\cr
&-|1+i\tilde{f}^i|\{|1+i\tilde{f}^i|\p_M|1+i\tilde{f}^k|\p_{\ov
N}|1+i\tilde{f}^k|+\cr
&+\left.|1+i\tilde{f}^k|(|1+i\tilde{f}^k|\p_M\p_{\ov
N}|1+i\tilde{f}^i|-|1+i\tilde{f}^i|\p_M\p_{\ov N}|1+i\tilde{f}^k|)\}\right],\ (M,N)\neq (T^k, T^k),\cr
R^{7,j}_{T^k \ov T^k i\ov{i}}&={-1\over (T^k-\ov T^k)^2}G_{C^{7,j}_i\ov C^{7,j}_i}-{\kappa_4^{-2}\over{(T^k-\ov{T}^k)(U^i-\ov
U^i)}}{1\over{|1+i\tilde{f}^i|^3|1+i\tilde{f}^k|}}\times\cr
&\times\left[|1+i\tilde{f}^k|^2\p_M|1+i\tilde{f}^i|\p_{\ov N}|1+i\tilde{f}^i|-\right.\cr
&-|1+i\tilde{f}^i|\{|1+i\tilde{f}^i|\p_M|1+i\tilde{f}^k|\p_{\ov
N}|1+i\tilde{f}^k|+\cr
&+\left.|1+i\tilde{f}^k|(|1+i\tilde{f}^k|\p_M\p_{\ov
N}|1+i\tilde{f}^i|-|1+i\tilde{f}^i|\p_M\p_{\ov N}|1+i\tilde{f}^k|)\}\right],
}}
\eqn\RmixedII{\eqalign{
R^{7,j}_{M\ov Nj\ov{j}}&={-\kappa_4^{-2}\over{(S-\ov S)(U^j-\ov{U}^j)}}\left[
\p_M\p_{\ov N}|1-\tilde f^i\tilde f^k|-\p_M|1-\tilde f^i\tilde f^k|\p_{\ov N}
|1-\tilde f^i\tilde f^k|{1\over|1-\tilde f^i\tilde f^k|}\right],\cr
&\quad(M,N)\neq(S,S),\cr
R^{7,j}_{S\ov Sj\ov{j}}&={-1\over (S-\ov S)^2}G_{C^{7,j}_j\ov C^{7,j}_j}-{\kappa_4^{-2}\over{(S-\ov S)(U^j-\ov{U}^j)}}\left[
\p_M\p_{\ov N}|1-\tilde f^i\tilde f^k|\right.\cr
&\left.-\p_M|1-\tilde f^i\tilde f^k|\p_{\ov N}
|1-\tilde f^i\tilde f^k|{1\over|1-\tilde f^i\tilde f^k|}\right],\cr
R^{7,j}_{M \ov U^ii\ov{i}}&=R^{7,j}_{M \ov U^jj\ov{j}}=0\ ,}}
where
\eqn\ugly{\eqalign{
\partial_M|1+i\tilde f^k|&=-|1+i\tilde f^k|^{-1}{(f^k)^2\over ({\rm Im}{\cal
T}^k)^3}\partial_M({\rm Im}{\cal T}^k),\cr
\partial_M\partial_{\ov N}|1+i\tilde f^k|&=|1+i\tilde f^k|^{-1}{(f^k)^2\over
({\rm Im}{\cal T}^k)^4[(f^k)^2+({\rm Im}{\cal
T}^k)^2]}\{[2(f^k)^2+3({\rm Im}{\cal T}^k)^2]\times\cr
&\times\partial_M({\rm Im}{\cal T}^k) \partial_{\ov N}
({\rm Im}{\cal T}^k)-{\rm Im}{\cal T}^k[(f^k)^2+({\rm Im}{\cal
T}^k)^2]\partial_M\partial_{\ov N}({\rm Im}{\cal T}^k)\},\cr
\p_M|1-\tilde f^i\tilde f^k|&={-f^if^k\over ({\rm Im}{\cal T}^i)^2({\rm Im}{\cal
T}^k)^2} \left[({\rm Im}{\cal T}^k)\p_M({\rm Im}{\cal T}^i)+({\rm Im}{\cal
T}^i)\p_M({\rm Im}{\cal T}^k)\right],\cr
\p_M\p_{\ov N}|1-\tilde f^i\tilde f^k|&={-f^if^k\over ({\rm Im}{\cal
T}^i)^3({\rm Im}{\cal T}^k)^3}\times\cr
&\times\left[
-2 ({\rm Im}{\cal T}^k)^2\p_M({\rm Im}{\cal T}^i) \p_{\ov N}({\rm Im}{\cal
T}^i)-
({\rm Im}{\cal T}^k)({\rm Im}{\cal T}^i)\p_{\ov N}({\rm Im}{\cal
T}^i)\p_M({\rm Im}{\cal T}^k)\right.\cr
&-({\rm Im}{\cal T}^k)({\rm Im}{\cal T}^i)\p_M({\rm Im}{\cal T}^i)\p_{\ov
N}({\rm Im}{\cal T}^k)- 2({\rm Im}{\cal T}^i)^2\p_M({\rm Im}{\cal
T}^k)\p_{\ov N}({\rm Im}{\cal T}^k)\cr
&+\left.({\rm Im}{\cal T}^i)({\rm Im}{\cal T}^k)^2\p_M\p_{\ov N}({\rm
Im}{\cal T}^i) +({\rm Im}{\cal T}^i)^2({\rm Im}{\cal T}^k)\p_M\p_{\ov N}({\rm Im}{\cal T}^k)
\right]\ .}}

Now we come to the terms coming from the twisted
matter metrics. From the metric for matter fields going from the 3-branes to
the $a$'th of the 7-brane stacks, we get the following components. Again,
$R^{37,a}_{U^j\ov U^j i \ov i}$ has the simplest structure and again,
$R^{37,a}_{U^j\ov U^k i \ov i},\ k\neq j$ is zero.
\eqn\twistedthree{\eqalign{
R^{37,a}_{U^j\ov U^j}&={-\theta^j \over {(U^j-\ov
U^j)^2}}G_{C^{37_a}\ov C^{37_a}},\quad j\neq a \cr
R^{37,j}_{U^j\ov U^j}&=0,\cr
R^{37,a}_{M\ov U^j}&={\partial_M\theta^j \over {(U^j-\ov
U^j)}}G_{C^{37_a}\ov C^{37_a}},\quad a\neq j,\cr
R^{37,j}_{M\ov U^j}&=0,\cr
R^{37,a}_{M\ov N}&=-\sum_{j\neq a}\ln(U^j-\ov
U^j)\ \partial_M\partial_{\ov N}(\theta^j)G_{C^{37_a}\ov C^{37_a}}\ ,}}
where 
\eqn\dtheta{\eqalign{
\partial_M\theta^j&=-{1\over \pi}{1\over {(1+(\tilde f^j)^2)}}{f^j\over {({\rm
Im}{\cal T}^j})^2}\ \partial_M({\rm Im}{\cal T}^j),\cr
\partial_M\partial_{\ov N}(\theta^j)&=-{f^j\over \pi\ [(f^j)^2+({\rm Im}{\cal
T}^j)^2]^2}\ \left[-2({\rm Im}{\cal T}^j)\ \partial_M({\rm Im}{\cal T}^j)
\partial_{\ov N}({\rm Im}{\cal T}^j)+\right.\cr
&\left.+[(f^j)^2+({\rm Im}{\cal
T}^j)^2]\ \partial_M\partial_{\ov N}({\rm Im}{\cal T}^j)\right].
}}
From the metric for matter fields going from the $a$'th 7-brane stack to the $b$'th
7-brane stack, we get
\eqn\twistedseven{\eqalign{
R^{7a,7b}_{U^k\ov U^k}&={-\theta^k_{ab} \over {(U^k-\ov
U^k)^2}}G_{C^{7a7b}\ov C^{7a7b}}  \cr
R^{7a,7b}_{U^j\ov U^k}&=0,\quad j\neq k,\cr
R^{7a,7b}_{M\ov U^k}&={\partial_M(\theta^k_{ab}) \over {(U^k-\ov
U^k)}}G_{C^{7a7b}\ov C^{7a7b}},\cr 
R^{7a,7b}_{M\ov N}&=\sum_l\left\{-\ln(U^l-\ov U^l)\partial_M\partial_{\ov
N}(\theta^l_{ab})\right.\cr
&+\half\left[\partial_M\partial_{\ov N}(\theta^l_{ab})[\psi_0(\theta^l_{ab})+
\psi_0(1-\theta^l_{ab})]+\right.\cr
&\left.\left.+\partial_M(\theta^l_{ab})\partial_{\ov
N}(\theta^l_{ab})[\psi_1(\theta^l_{ab})-\psi_1(1-\theta^l_{ab})]\right]\right\}G_{C^{7a7b}\ov C^{7a7b}},
}}
where $\psi_n$ is the $n$'th Polygamma function and
\eqn\dthetaab{\eqalign{
\partial_M\theta^j_{ab}&=-{1\over \pi}\left[{1\over {1+(\tilde
f_b^j)^2}}{f_b^j\over  ({\rm Im}\ {\cal T}^j)^2}\ \partial_M({\rm Im}{\cal T}^j)
-{1\over {1+(\tilde f_a^j)^2}}{f_a^j\over({\rm Im}\ {\cal T}^j)^2}\partial_M 
({\rm Im}{\cal T}^j)\right],\cr
\ \partial_M\partial_{\ov N}(\theta^j_{ab})&=-{1\over \pi}
\left\{{f^j_b\over[(f_b^j)^2+({\rm Im}\ {\cal T}^j)^2]^2} \left[-2({\rm Im}
{\cal T}^j)\partial_M ({\rm Im}{\cal T}^j) 
\ \partial_{\ov N} ({\rm Im}{\cal T}^j)+\right.\right.\cr
&\left.+[(f_b^j)^2+({\rm Im}{\cal T}^j)^2]\ \partial_M\partial_{\ov N}({\rm Im}
{\cal T}^j)\right]-\cr
&-{f^j_a\over{[(f_a^j)^2+({\rm Im}{\cal T}^j)^2]}^2}\left[-2({\rm Im}
{\cal T}^j)\ \partial_M ({\rm Im}{\cal T}^j)
\ \partial_{\ov N}({\rm Im}{\cal T}^j)+\right.\cr
&\left.\left.+[(f_a^j)^2+({\rm Im}{\cal T}^j)^2]\ \partial_M\partial_{\ov N}
({\rm Im}{\cal T}^j)\right]\right\}.
}}

\listrefs
\end